\documentclass{aa}  

\usepackage{graphicx}
\usepackage{txfonts}
\usepackage{xcolor}
\usepackage[pdfpagelabels=false]{hyperref}	% Hyperlinks
\hypersetup{colorlinks=true,linkcolor=blue,citecolor=blue,filecolor=blue,urlcolor=blue,}
\begin{document}

   \title{Measurement of circumgalactic extinction in the Kilo-Degree Survey's Data Release 4}
   \titlerunning{Circumgalactic extinction in KiDS DR4}

   %\subtitle{I. Overviewing the $\kappa$-mechanism}

   \author{Eray Genc
          \inst{1}\fnmsep\thanks{\email{egenc@astro.ruhr-uni-bochum.de}}
          \and 
          Angus H. Wright\inst{1} %\fnmsep\thanks{Just to show the usage
          %of the elements in the author field}
          \and 
          Hendrik Hildebrandt\inst{1}
          }

   \institute{Ruhr University Bochum, Faculty of Physics and Astronomy, Astronomical Institute (AIRUB), German Centre for Cosmological Lensing, 44780 Bochum, Germany       
         %\and
         %    Ruhr-Universität Bochum, Astronomisches Institut, German Centre for Cosmological Lensing (GCCL), Universitätsstr. 150, 44801 Bochum, Germany\\
            % \email{c.ptolemy@hipparch.uheaven.space}
             %\thanks{The university of heaven temporarily does not
            %         accept e-mails}
             }

   \date{Received XXX; accepted XXX}

% \abstract{}{}{}{}{} 
% 5 {} token are mandatory
 
   \abstract
  % context heading (optional)
  % {} leave it empty if necessary  
   {Galaxies reside in extended dark matter halos, but their baryonic content, especially dust, remains less well understood.}
  % aims heading (mandatory)
   {We aim to investigate the amount and distribution of dust in galactic halos by measuring shifts in the observed magnitudes of background galaxies.}
  % methods heading (mandatory)
   {Dust absorbs and scatters shorter-wavelength light, reddening background sources. We quantify this effect using multi-band magnitude shifts in the Kilo-Degree Survey (KiDS) Data Release 4, distinguishing it from achromatic magnification. Our pipeline is validated using MICE2 mock catalogues, and we explore potential systematics arising from galaxy selection, in contrast to previous quasar-based studies.}
  % results heading (mandatory)
   {Simulations show that input halo and dust masses can be recovered independently and jointly using galaxy-galaxy lensing (GGL) and magnitude shifts. A minimum redshift separation of $\Delta z_{\mathrm B}\sim0.3$ and brightness cuts are required to minimise overlap effects and the effects of photometric noise, respectively. Applying our pipeline to KiDS data, we first confirm the achromaticity of magnitude shifts in five near-infrared filters ($ZYJHK_{\rm s}$). In optical bands ($ugri$), joint analysis constrains halo mass and dust mass to $0.2$ dex and $0.3$ dex, respectively, yielding $M_{\rm dust}=7.24^{+0.52}_{-0.64} \times 10^7\,M_\odot$ and $\log(M_{\rm halo}^{\rm DM}/M_\odot)=12.33^{+0.08}_{-0.04}$. These results agree with previous studies and represent the first successful measurement of magnitude shifts and halo extinction with KiDS, confirming that galaxies reside in extended baryonic halos enriched by gas and dust from feedback processes.}
  % conclusions heading (optional), leave it empty if necessary 
  {}
   

\maketitle
%
%-------------------------------------------------------------------
\section{Introduction}
Understanding the distributions of dark and baryonic matter in galaxies is crucial for many facets of astronomical research, from simulations of galaxy formation to cosmological measurements with weak gravitational lensing. Using gravitational lensing measurements, such as galaxy-galaxy lensing (GGL), cosmic shear, and magnification, the statistical properties of dark matter in galaxy halos have become well understood. However, the same is less true for the baryonic components of galaxy halos, such as gas and dust. 

Dust grains are organic solids produced by stars and can be transported into the circumgalactic medium (CGM) via radiation pressure from stars \citep{bianchi_2005} or by galactic winds driven by supernovae \citep{ag_2001}. Knowledge of the amount, and extent, of dust in galaxy halos can provide crucial insights into the physical processes that govern galaxy formation and evolution, as dust is directly linked to star formation activity, feedback processes, and (therefore) galaxy morphology. Moreover, investigation of the distribution of dust in halos at different redshifts can provide constraints on the evolution of galaxies, as it traces feedback processes, and potentially encodes information about the metallicity and star formation history of galaxies. 

\citet{menard_2010} presented the first observational measurement of the amount, and distribution, of dust in galactic halos. In their analysis, the authors leveraged high-redshift quasars to measure lensing-induced shifts in the apparent brightness of background sources around foreground ‘lens’ galaxies in the Sloan Digital Sky Survey \citep[SDSS,][]{york_2000}. By measuring the apparent brightness shifts of background sources in various photometric bandpasses, the authors quantified the wavelength dependence of these shifts, which serves as a tracer of dust in the halos of the lens galaxies. They detected significant chromaticity in the apparent brightness shifts of quasars, on radial scales ranging from 20 kpc to a few Mpc, indicating the presence of significant quantities of dust in the halos of their lens galaxies.

Since the work by \citet{menard_2010}, the study of circumgalactic dust has seen relatively limited advancement. However, current imaging surveys for weak lensing, such as the Kilo-Degree Survey \citep[KiDS,][]{de2013,wright_2024}, present optimal datasets for performing such measurements. These datasets are observed over wide areas on-sky, feature high precision photometric measurements across a range of wavelengths, have high fidelity photometric redshift estimates, and exhibit well behaved selection functions; all of which are required to perform accurate and precise measurements of magnitude shift. 
%One of the recent works aiming to measure reddening by dust was conducted by \citet{bell_2019}. Authors quantified the distribution of dust grains in Small Magellanic Cloud (SMC) by examining the reddening in the spectral energy distributions (SEDs) of background galaxies. \citet{brout_2023} reviewed the potential impacts of intergalactic and interstellar dust on cosmic distance measurement methods. 

In this work, we aim to measure both the mean halo mass and the mean circumgalactic dust mass of a well-defined sample of galaxies from the fourth data release of KiDS (KiDS-DR4). Our approach builds on the methods employed by \citet{menard_2010}, with some key modifications. We utilise magnitude shift and galaxy-galaxy lensing signals for the estimation of mean halo mass, and utilise the chromaticity of the apparent brightness shifts of galaxies across four optical filters/bandpasses for the estimation of circumgalactic dust mass. Additionally, we use galaxies as our source population for all measurements, as opposed to quasars \citep{menard_2010}.

Prior to performing our measurements on data from KiDS-DR4, we validate our analysis methods and pipelines using simulations. For this purpose we use the MICE2 suite of simulations \citep{fosalba_2015a, crocce_2015, fosalba_2015b},  with catalogue properties modified by \citet{jlvdb_2020} to better match KiDS-DR4 data. These simulations are designed to mimic the observational properties of KiDS-DR4, and thus allow us to assess both the statistical properties of our measurements and explore any systematic biases inherent to the methodology. In particular, we pay attention to the development of robust selection criteria for background sources, in an effort to mitigate systematic effects that would otherwise require modelling and marginalisation. In the second part of this study, we apply our validated pipeline to the actual KiDS-DR4 dataset. 

Throughout this analysis, when the assumption of a cosmological model is required, we adopt a \citet{planck_2018} model. Our manuscript is structured as follows: the theoretical framework of our approach is presented in Sect. \ref{sec:2}; we describe the KiDS-DR4 data set in Sect. \ref{sec:3}; we present simulation results from MICE2 in Sect. \ref{sec:4}, and the first measurements of circumgalactic dust with KiDS-DR4 in Sect. \ref{sec:5}.

%--------------------------------------------------------------------
\section{Theoretical framework}
\label{sec:2}
In the following we assume a simple geometry consisting of two populations of galaxies along the line of sight. Foreground galaxies (`lenses') deflect the light rays propagating from background galaxies (`sources') and cause a magnification of their apparent image (relative to their unlensed equivalent). This magnification, which is caused by the gravitational potential of the foreground lens, increases in the apparent size of the source galaxy image while conserving surface brightness, thereby resulting in a net increase in the apparent total flux of the source. The magnitude of this magnification can therefore be expressed as the ratio of observed solid angle $\omega$ of a source galaxy's image to its intrinsic image size $\omega_0$ (i.e. that which we would observe in the absence of magnification). This can be calculated from the Jacobian matrix of the gravitational potential $\mathcal{A}$ \citep{bartelmann_2001}\footnote{The Jacobian matrix $\mathcal{A}$ is the linear mapping of source plane onto the lens plane describing the distortion of source images.}:
\begin{equation}
    \mu = \frac{\omega}{\omega_0} = \frac{1}{\rm{det}\mathcal{A}} = 
    \frac{1}{(1-\kappa)^2-|\gamma|^2}\;,
\end{equation}
where $\kappa$ is the dimensionless lensing convergence and $\gamma$ is the shear, which together quantify the distortion of light rays due to the gravitational potential of foreground objects. 
In the weak-lensing limit, magnification can solely be expressed as a function of $\kappa$:
\begin{equation}
    \mu \approx 1+2\kappa\;.
\end{equation}
Furthermore, $\kappa$ is equal to the ratio of the surface mass density of the lens galaxy, $\Sigma$, to the critical surface mass density:
\begin{equation}
 \Sigma_{\rm crit} =\frac{c^2}{4\pi G}\frac{D_{\rm s}}{D_{\rm d}D_{\rm ds}}\;,
\end{equation}
where $D_{\rm s}$, $D_{\rm d}$, and $D_{\rm ds}$ are the angular diameter distances to the source and lens planes, and the distance between them, respectively.
%This gives
%\begin{equation}
%    \mu \approx 1+\frac{8\pi G}{c^2}\frac{D_{\rm d}D_{\rm ds}}{{D_{\rm s}}}\Sigma. 
%\end{equation}
%

\subsection{Measurement of magnitude shift}\label{sec:magmeas}
We measure the shift in apparent magnitudes of galaxies by computing the cross-correlation between the number density of foreground galaxies and the brightness of background galaxies:
\begin{equation}
    \langle \delta m_{\rm obs}\rangle(\theta) \simeq \langle \delta_{\rm g}(\phi+\theta)\,\Delta m(\phi) \rangle, 
\end{equation}
where $\delta m_{\rm obs}$ is the observed magnitude shift of the source sample, $\delta_{\rm g}(\phi+\theta)$ describes the galaxy overdensity in the foreground, and $\Delta m(\phi) = m(\phi) - \langle m\rangle$ is the residual brightness of the sources. Here, $\phi$ denotes the angular position of the sources on the sky, while $\theta$ is the angular separation between the sources and the lenses.

The above formalism assumes that all sources, regardless of their pre- and post-magnification brightness, are able to be measured. In practice, though, imaging surveys for cosmology are typically magnitude limited in one or more photometric bands. Magnification can cause sources which are intrinsically fainter than the limit to be brightened such that they move above the limit and are therefore detected. If the source galaxy number counts rise sharply at the magnitude limit, this can result in a net increase in the number of intrinsically faint sources appearing within the detection window, which can skew the measured magnitude offset. To account for this possible bias, \citet{menard_2010} introduced the so-called $C_\mathrm{s}$ coefficient, which scales the intrinsic magnification signal:  
\begin{equation}
    \delta m_{\rm obs} \simeq C_\mathrm{s} \times \delta m_{\rm int}.
\end{equation}
As the expected bias is dependent on the slope of the source number counts $N$, $C_\mathrm{s}$ is calculated from the differential source number counts as a function of magnitude $\mathrm{d}N/\mathrm{d}m$: 
\begin{equation}
    C_\mathrm{s} = 1 - \frac{1}{N_{0,\rm tot}}\frac{\mathrm{d}N}{\mathrm{d}m}(m_{\rm faint})\,[m_{\rm faint} - \langle m_0\rangle],
\end{equation}
where $N_{0,\rm tot}$ and $\langle m_0\rangle$ are the total number of galaxies and the average magnitude of the source sample after the brightness cut $m_{\rm faint}$, respectively. The subscript $0$ indicates that these values are intrinsic, that is as measured in absence of magnification. As this is a fundamentally unknowable quantity, we assume they are equal to the observed values for a sufficiently large sample of sources sampled over the full survey.
Therefore, the induced magnification can be evaluated from our observables as
\begin{equation}
    \langle \delta_{\rm g}(\phi+\theta)\,\delta\mu(\phi) \rangle \simeq \frac{\langle \delta_{\rm g}(\phi+\theta)\,\Delta m(\phi) \rangle}{C_\mathrm{s}},
\end{equation}
where $\delta\mu\approx 1-\mu$ is the relative magnification, which is used as it is more convenient in magnification measurements in the weak-lensing limit. It should be noted, though, that this measurement is expected to overestimate the true magnification signal on angular scales below a few arcminutes \citep{menard_2003}, where non-linear effects on lensing signals become significant and the weak-lensing approximation begins to break down.

Given our above estimates of magnification, we can subsequently compute the mean surface mass density $\langle\Sigma_{\mu}\rangle(\theta)$ as inferred from our magnification signal, through a straight multiplication with the critical surface mass density:
\begin{equation}
    \langle\Sigma_{\mu}\rangle(\theta) = \Sigma_{\rm crit}\,\langle\delta_{\rm g}(\phi+\theta)\,\delta\mu(\phi)\rangle\;.
\end{equation}

\subsection{Reddening measurements}
The formalism of our magnitude shift measurements has not required any consideration of the wavelength of bandpass that has been used for the measurement, as the magnification process is achromatic: the increase in brightness for a source is not wavelength dependent. However, as previously discussed, dust in the halos of foreground galaxies preferentially absorbs and scatters light rays at shorter wavelengths. As such, light travelling through the halos of lens galaxies, in the presence of dust, ought to experience a wavelength-dependent extinction, known typically as reddening. By quantifying the magnitude of this reddening, we can infer information about the amount and distribution of circumgalactic dust in halos of our lens galaxies. 

In Sect.~\ref{sec:magmeas} we introduced the magnitude change of background galaxies with respect to the mean magnitude of the source sample. If such a measurement is made in an arbitrary photometric filter $x$, we have
\begin{equation}
    \delta m_x(\phi) = m_x(\phi) - \langle m_x \rangle.
\end{equation}
We can subsequently define the colour excess between magnitude shift measurements made in two specific photometric filters $x=p$ and $x=q$, to quantify the reddening due to circumgalactic dust, as
\begin{equation}
    E_{pq} = E(\lambda_p - \lambda_q) = \delta m_p - \delta m_q\;.
\end{equation}
We can then define the density-colour excess correlation function as
\begin{equation}
    \langle \delta_{\rm g}(\phi+\theta)\,E_{pq}(\phi)\rangle \simeq 1.08\,\langle [\tau(\lambda_p) - \tau(\lambda_q)]\rangle (\theta),
\end{equation}
where $\tau(\lambda_x)$ is the optical depth of dust in filter $x$. The factor 1.08 arises from the conversion between magnitudes and optical depth via the relation:
\begin{equation}
    \delta m = 2.5\,\mathrm{log}_{10}(F_{\rm obs}/F_{\rm int})=2.5\mathrm{log}_{10}(e)\,\tau\sim1.08\,\tau
\end{equation}
Here, $F_{\rm obs}$ and $F_{\rm int}$ are the observed and intrinsic fluxes of a background source, respectively, and extinction is assumed to follow an exponential attenuation law $F_{\rm obs} = F_{\rm int}\times e^{-\tau}$.
So, by measuring the mean colour excess of background galaxies, as a function of angular separation to the foreground galaxy positions, we can estimate the mean optical depth of dust in the CGM of foreground lens galaxies. 

Finally, we can estimate the surface mass density profile of dust $\Sigma_{\rm dust}$ within galactic halos using our reddening measurements, which we calculate by dividing the mean optical depth of dust by the absorption optical depth per unit dust mass $K_{\rm ext}$:
\begin{equation}
  \Sigma_{\rm dust} = \frac{\tau_{V}(r_p)}{K_{\rm ext}(\lambda_V)}\simeq \frac{E(B-V)}{1.08\times K_{\rm ext}(\lambda_V)}\approx\frac{E(g-i)}{1.08\times K_{\rm ext}(\lambda_V)}.
  \label{eq:sd}
\end{equation}
Consistent with \citet{menard_2010}, we assume $E(B-V)\approx E(g-i)$ and adopt a Small Magellanic Cloud (SMC) type dust model proposed by \citet{w_draine_2001}, which yields $K_{\rm ext}=3.217\,{\rm pc}^2{M_\odot}^{-1}$.
To quantify the mean dust mass in the halos of our lens galaxies, we then simply integrate the surface density of dust mass as a function of separation:
\begin{equation}
    M_{\rm dust}^{\rm halo} = \frac{2\pi\,\mathrm{ln}10}{2.5\,K_{\rm ext}(\lambda_V)} \int_{r_\mathrm{l}(z)}^{r_\mathrm{u}(z)} A_V(r_p)\,r_p\,dr,
    \label{eq:dust_mass}
\end{equation}
where $A_V(r_p)$ is the absorption profile in $V$-band. We select the lower bound of this integral to avoid contributions from dust contained within the interstellar medium of our lenses (typically using $r_\mathrm{l}(z)=20\,h^{-1}{\rm kpc}$). For the upper bound, we use the approximate virial radius of a galactic halo, measured at the redshift of the lens:
\begin{equation}
    R_{\rm vir}(z) = \left( \frac{M_{\rm halo}^{\rm DM}}{\Delta_c(z)\, \rho_{\rm crit}(z)} \right)^{1/3}\;.
\end{equation}
Here, $M_{\rm halo}^{\rm DM}$ denotes the halo mass, $\Delta_c(z)$ is the overdensity parameter, and the $\rho_{\rm crit}$ is the critical density of the Universe at redshift $z$.

\subsection{Galaxy-galaxy lensing}
As the light rays coming from distant sources are distorted by the mass distribution of foreground galaxies, there is a correlation between the shapes of sources and positions of lens galaxies. By measuring the mean tangential shear of the sources as a function of their separation from the lenses, one can infer information about the mass distribution of the lens galaxies:
\begin{equation}
    \langle\gamma_{\rm t}\rangle(r) = \frac{\Delta\Sigma (R)}{\Sigma_{\rm crit}},
\end{equation}
where $\Delta\Sigma (R)$ is the excess surface mass density within an angular or physical radius $R$.

\subsection{Estimation of halo parameters}
To estimate the mean halo mass from the above, we employ a theoretical model which leverages an assumed density profile for the distribution of matter in our lenses. We use the Navarro-Frenk-White (NFW) density profile \citep{nfw_97}:
\begin{equation}
    \rho(R) = \frac{\rho_\mathrm{s}}{\left( \frac{R}{r_\mathrm{s}} \right) \left( 1 + \frac{R}{r_\mathrm{s}} \right)^2}\;,
\end{equation}
where $\rho(r)$ is the density as a function of physical separation $R$, $\rho_s$ denotes the characteristic density and $r_\mathrm{s}$ is the scale radius.
While the NFW profile itself is defined by the parameters $r_\mathrm{s}$ and $\rho_s$ in our modeling framework these are derived from the halo mass and redshift of the lens via a mass–concentration relation \citep{diemer_2019}. This allows us to express the density profile in terms of physically meaningful quantities such as halo mass. From this model, we derive two key quantities: the surface mass density contrast $\Delta \Sigma(R)$, which we can compare with our measured GGL signal, and the surface mass density $\Sigma(R)$ that we can compare with our measured magnitude shift signals. We can therefore perform a straightforward optimisation of our model to fit our observed signals, either independently or jointly, and determine the best-fitting parameters that describe the underlying mass distribution of our lens halos.

We further develop our density profile model to account for the so-called two-halo contribution. On large scales one also expects a contribution to the observed signal from other halos (i.e. not the host halo of the lens). This effect is commonly modelled using the linear matter power spectrum combined with a halo bias prescription. For our purposes, we adopt an analytic approximation to this contribution, referred to as the infalling profile, which has been shown to capture the two-halo term behaviour with sufficient accuracy for our measurement precision \citep{diemer_2023}.

We use the implementation provided by the Colossus code \citep{diemer_2018}, assuming a fixed flat $\Lambda$CDM cosmology consistent with \citet{planck_2018} results. Colossus internally computes the linear matter power spectrum and derives the large-scale bias of halos based on the fitting function from \citet{tinker_2010}, which expresses the halo bias as a function of halo mass and redshift. The resulting infalling profile is expressed as:
\begin{equation}
    \label{eq:rho}
    \rho(R) = \delta_1\rho_{\rm m}(z)\, \left[\left(\frac{\delta_1}{\delta_{\rm max}}\right)^{2} + \left(\frac{R}{r_{\rm pivot}}\right)^{s/0.5}\right]^{-0.5},
\end{equation}
where $\delta_1$ is the normalization of the profile at the pivot radius $R_{200}$ and $s$ is the slope of the profile.
%---------------------------------------------------------------------
\section{Data}
\label{sec:3}
In this manuscript we perform two principal analyses: a measurement on bespoke simulations, as a validation of our methods and pipelines, and a measurement on weak lensing imaging data from KiDS. We test our measurement pipeline using simulations as there the true (input) signals are known, and we can directly compare the accuracy and precision of our measurements and methods under various analysis and modelling choices. Additionally, by using realistic simulated catalogues that are designed to mimic our KiDS dataset, we can directly investigate the influence of various systematic and observational effects on our measurements. To that end, we utilise pre-existing mock galaxy catalogues from \citet{jlvdb_2020}\footnote{\url{https://github.com/KiDS-WL/MICE2_mocks}}, which were constructed for redshift distribution calibration testing in KiDS DR4. This mock catalogue is built on the MICE2 suite of simulations. In Sect.~\ref{subsec:3.1} we provide a detailed description of the MICE2 mock galaxy sample, and in Sect.~\ref{subsec:3.2} we describe our observational dataset from KiDS-DR4.

\subsection{MICE2 Simulation}
\label{subsec:3.1}
MICE2 is a dark matter only $N-$body simulation, which tracks the linear and non-linear growth of structures using $\sim$70 billion particles in a cube with a side-length of $3h^{-1}\,\mathrm{Gpc}$ \citep{fosalba_2015a}. This structure growth is tracked within the framework of a flat $\Lambda$CDM cosmological model with parameters $\Omega_\mathrm{m}=0.25$, $\Omega_\Lambda=0.75$, and $h=0.7$, from $z=100$ to the present.

The resulting halo catalogue is populated with galaxies using both halo occupation distribution (HOD) and subhalo abundance matching (SHAM) methods \citep{Carretero_2015}. The galaxy catalogue contains a number of fundamental observables, such as galaxy positions, redshifts, intrinsic magnitudes in a range of filters, and lensing quantities such as shear and convergence. Lensing observables are derived from an on-sky pixel grid with a resolution of $0\farcm43$, which limits the analysis of gravitational lensing effects to angular scales larger than this resolution. In the case of GGL, the finite resolution introduces a smoothing effect that is particularly impactful at small scales, as shear is highly sensitive to local variations in the lensing potential. The directional information of background galaxy shapes within a single pixel can partially or fully cancel out, especially when multiple galaxies with different intrinsic orientations fall into the same pixel. This leads to a suppression of the signal on angular scales comparable to or smaller than the pixel size. In contrast, magnification is a scalar (spin-0) quantity and is intrinsically less sensitive to small-scale fluctuations in the lensing potential. As a result, magnification measurements are expected to be less affected by the finite resolution of the lensing map, allowing for reliable extraction of the signal down to smaller angular scales compared to GGL.

Additionally, the simulation is designed to reproduce key observational distributions, such as the luminosity function and colour-magnitude distribution observed by the SDSS. Therefore,  the simulation provides a realistic representation of galaxy populations in the Universe, and is well-suited for use in our analysis. 

The KiDS-like mock catalogue developed by \citet{jlvdb_2020} uses SDSS filters for the magnitude estimates in  the $ugriz$ bands, the Dark Energy Survey \citep[DES, ][]{flaugher_2015} $Y$ band, and the VIKING \citep[VISTA Kilo-degree Infrared Galaxy Survey, ][]{edge_2013} filters for $JHK_{\mathrm{s}}$ bands. In the catalogue, three types of magnitudes are provided: intrinsic magnitudes including evolutionary corrections ($m_{\rm evo}$), additional intrinsic magnitudes that also account for magnification ($m_{\rm mag}$), and realistic KiDS- and VIKING-like magnitudes including photometric noise ($m_\mathrm{obs}$). 
The MICE2 evolutionary correction is necessary for agreement between simulated and observed galaxy number counts, and is defined by \citet{fosalba_2015b} as:
\begin{equation}
    m_{\rm evo} = m_0 - 0.8\,[\mathrm{arctan}(1.5\,z_{\rm cgal})-0.1489],
\end{equation}
where $m_0$ is the intrinsic galaxy magnitude produced by the MICE2 semi-analytic modelling, and $z_{\rm cgal}$ is the galaxy cosmological redshift (i.e. excluding the peculiar velocity component of the observed redshift).
The magnification correction to the magnitudes is defined using the weak-lensing approximation ($\delta\mu\approx2\kappa$):
\begin{equation}
    m_{\rm mag} = m_{\rm evo} - 2.5\,\mathrm{log}_{10}(1+\delta\mu),
\end{equation}
and is able to be optionally included, thereby allowing us to perform both realistic analyses and null-tests with our simulated galaxy catalogues. These magnitude definitions provide a flexible framework for simulating various lensing scenarios. However, they do not include circumgalactic dust extinction, therefore, we implement this effect in our mock data, as described in the following section.

\subsubsection{Simulation of dust extinction}
Extinction due to circumgalactic dust is not present in MICE2, nor the KiDS-like catalogue by \citet{jlvdb_2020}. Therefore, we adjust the galaxy magnitudes in our sample following the recommendations of \citet{garcia_2018}. Specifically, we apply the modification to the galaxy magnitudes: 
\begin{equation}
\label{eq:redd}
    m_{\rm ext} = m_{\rm mag} + 1.09\,\tau_{\rm V}\,\frac{\lambda_{\rm V}}{\lambda}\sum_{l}\left(\frac{r_p}{\beta}\right)^{-\alpha},
\end{equation}
where $\alpha$ and $\beta$ are tuning parameters that adjust the shape and amplitude of the simulated extinction profile, while $l$ represents each lens galaxy, as we sum up the extinction contributions from foreground galaxies for each source galaxy.

\begin{figure}
    \centering
    \includegraphics[width=\columnwidth]{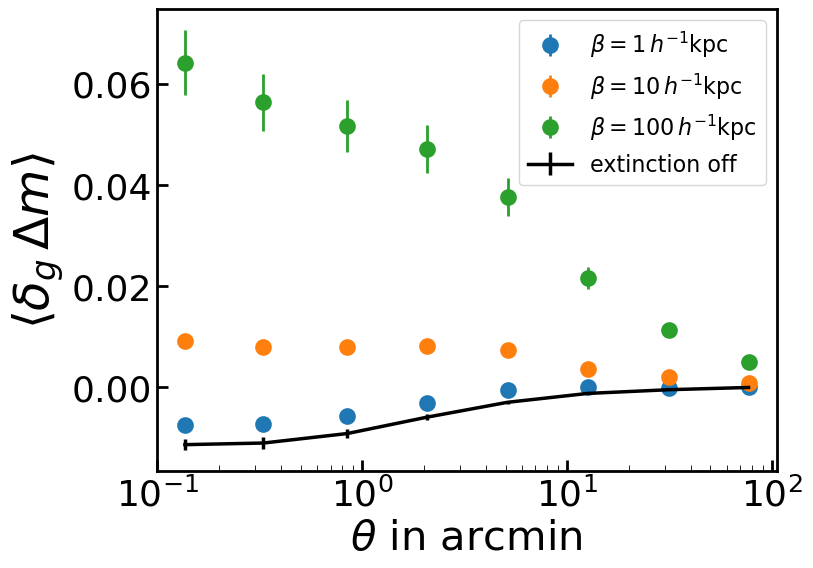}
    \caption{Magnitude shift measurements with simulated extinction profiles with different values for the parameters $\beta$. Parameter $\alpha$ is fixed to $0.8$, and extinction profiles are simulated with $\beta=1,\,10,\,100\,h^{-1}$ kpc. We do not apply any stellar mass cut to the lens galaxies.}
    \label{fig:simdust_alpha}
\end{figure}
\begin{figure}
   \centering
    \includegraphics[width=\columnwidth]{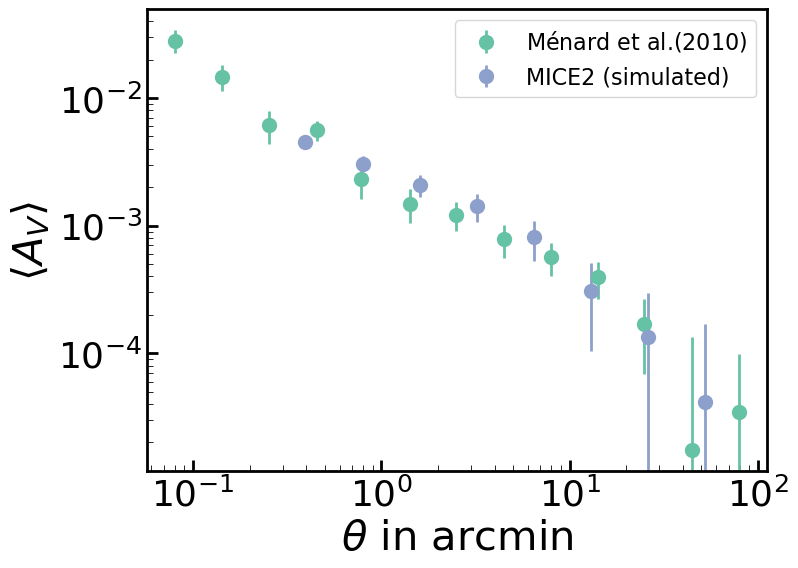}
    \caption{Comparison of the $V$-band extinction profile measured by \citet{menard_2010} with the one implemented in our simulation. For consistency, we selected MICE2 galaxies with a matching mean redshift. The close agreement indicates that our simulated extinction profile is consistent with observational results.}
    \label{fig:menard_ext_compare}
\end{figure}
This approach is grounded in the observational results from \citet{menard_2010}, and for a more realistic approach, we account for the projected physical separation in the lens plane, rather than relying on angular separations. We adopt the parameter values for $\alpha$ and $\beta$ from \citet{menard_2010}, specifically $\alpha = 0.8$ and $\beta = 100\, h^{-1}$ kpc.

To ensure a consistent comparison with the observational analysis by \citet{menard_2010}, we restrict dust extinction in our simulations to originate only from galaxies with stellar mass $\log(M_*/M_\odot) > 10.2$, matching the mean stellar mass of their lens sample. This cut is also physically motivated: since extinction accumulates along the line of sight, including the contribution of numerous low-mass galaxies without such a threshold leads to an overestimation of the total extinction. As shown in Fig.~\ref{fig:simdust_alpha}, omitting the mass cut causes extinction to dominate over the magnification signal, suppressing it to unrealistic levels and resulting in a strong mismatch with observational expectations. In Fig.~\ref{fig:menard_ext_compare}, we compare our simulated $V$-band extinction profile to that measured by \citet{menard_2010} for galaxies at similar redshift ($z \sim 0.3$). The agreement in both slope and amplitude validates our model implementation. We note that, in both figures and during the modification of galaxy magnitudes in the catalogue, we employed the true redshift information rather than the photometric redshift. Furthermore, while generating the figures, we switched off the photometric noise in the magnitudes in order to isolate the effect of extinction on the measured magnitude shifts.

\begin{figure*}
    \centering
    \includegraphics[width=0.47\textwidth]{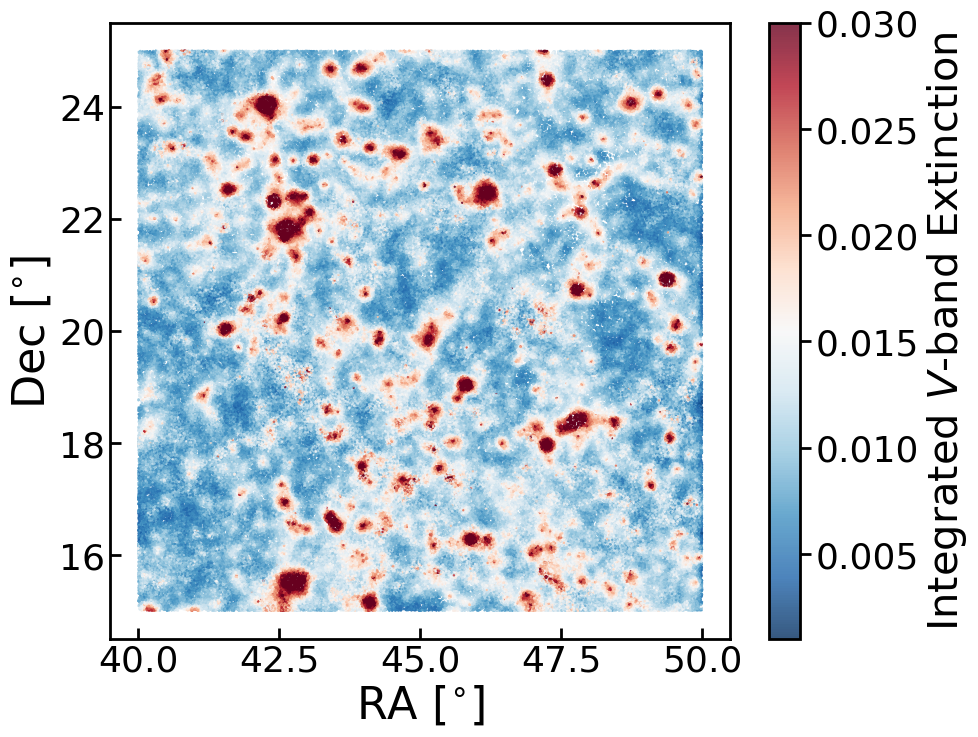}\hspace{1mm}
    \includegraphics[width=0.37\textwidth]{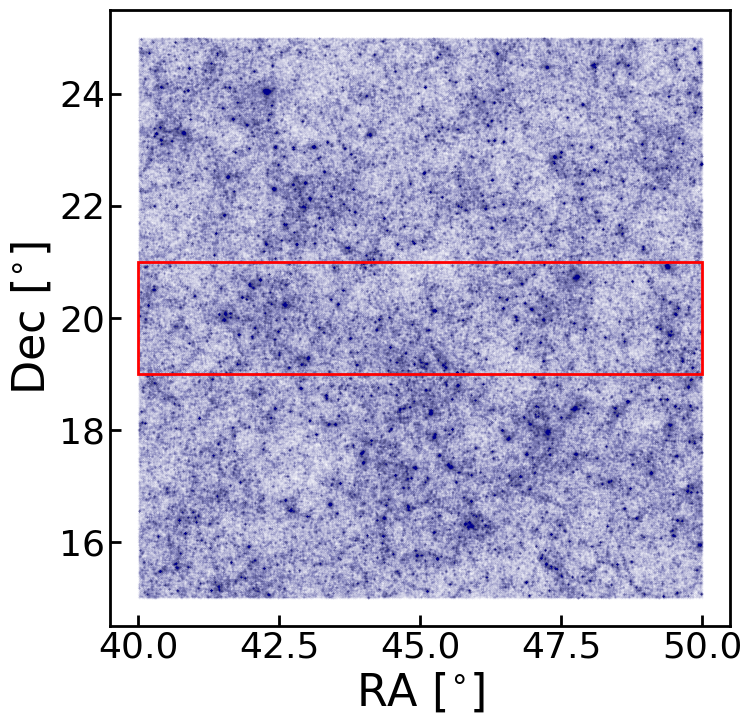}\\
    \includegraphics[width=0.6\textwidth]{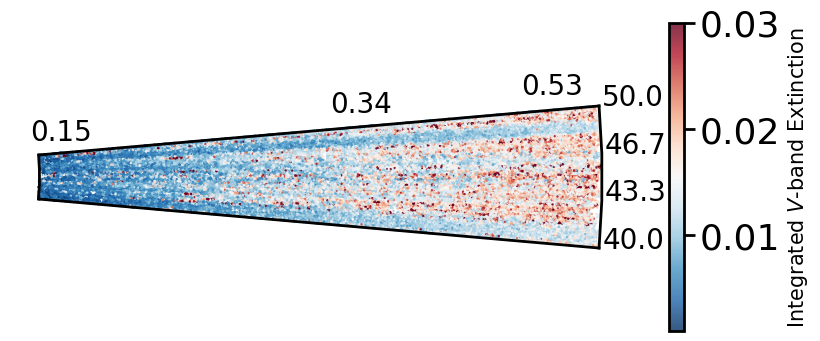}
    \caption{\textit{Upper left}: The simulated extinction in the $V$-band for galaxies within the redshift range $0.25 < z < 0.35$, integrated along the line of sight.
    \textit{Upper right}: The large-scale structure in the same region, but over a wider redshift range $0.15 < z < 0.52$. The red box highlights the specific region displayed in the lower panel. \textit{Lower panel}: The integrated $V$-band extinction in RA-$z$ space, focussing on sources within a narrow declination slice of $19^\circ < \mathrm{Dec} < 21^\circ$. These visualizations illustrate the spatial distribution and redshift evolution of the simulated $V$-band extinction, providing a qualitative check of the implemented dust model and its correlation with the underlying large-scale structure.}
    \label{fig:sim_results}
\end{figure*}

Additionally, Fig.~\ref{fig:sim_results} presents the integrated extinction map in the $V$-band generated from our simulation. As expected, the extinction shows a cumulative trend with increasing redshift, since we account for the contribution of each intervening halo along the line of sight to every source galaxy. The resulting extinction values range from 0 to 0.03 mag, corresponding to roughly $10\%$ of the typical Galactic extinction in the $r$-band (0.02–0.3 mag). For context, these simulated extinction values also lie within the reported zero-point uncertainties of KiDS photometry \citep{wright_2024}. However, it is important to note that while KiDS zero-point errors are uncorrelated across the sky, the extinction induced by CGM dust is spatially correlated.

\subsection{Kilo-Degree Survey}
\label{subsec:3.2}
KiDS is a wide-field imaging survey for cosmology, conducted on the 2.6m VLT Survey Telescope \citep[VST, ][]{capaccioli_2005} at the European Southern Observatory's (ESO) Paranal site in Chile. Observations with the VST are performed with the OmegaCAM wide-field camera \citep{kuijken_2011}, which has a field of view of $1^\circ\times1^\circ$ with a pixel scale of $0\farcs213$ per pixel. KiDS observed $\sim\!\!1345$ $\mathrm{deg}^2$ of the sky in the northern and southern Galactic caps, using four optical bands ($ugri$). We use data from the fourth data release of KiDS \citep[DR4, ][]{kuijken_2019}, which covers approximately $1000$ $\mathrm{deg}^2$ and provides various useful data products, including $r$-band detection images and multi-band source catalogues.
In addition to the four optical bands, the KiDS dataset also provides galaxy photometry in five near-infrared (NIR) bands ($ZYJHK_\mathrm{s}$), drawn from imaging taken by the VIKING survey \citep{edge_2013}, which was conducted on the 4m Visible and Infrared Survey Telescope for Astronomy (VISTA) also located at ESO's Cerro Paranal observatory in Chile. Observations were carried out using the Visible and Infrared Camera (VIRCAM), in the same on-sky footprint as KiDS. 
Matched aperture photometry across all filters $ugriZYJHK_{\mathrm{s}}$ is measured using the Gaussian aperture and point-spread function \citep[\texttt{GAaP}][]{kuijken_2015} code, which is optimised for accurate colours, using source detection and aperture definition based in the $r$-band. 

The KiDS dataset represents the ideal sample to use for our analysis, as the weak-lensing focus of the survey necessitates high fidelity photometric measurements in multiple bands from the ultra-violet to the near-infrared. 
As such, in our analysis, we make use of the KiDS-1000 weak lensing SOM-gold catalogue \citep{hildebrandt2021kids,asgari2021} for source galaxies, which contains approximately 21 million sources with robust photometric, shape, and photometric redshift ($z_B$ estimates. Photometric redshifts are estimated using the Bayesian photometric redshift code \citep[\texttt{BPZ}, ][]{benitez}, which employs a template-fitting approach with Bayesian priors that depend on the galaxy type and magnitude. 

Additionally, the KiDS team also provide high precision photometric redshift estimates for a sample of bright sources, defined as having $r$-band magnitudes brighter than 20. This so-called `bright sample' \citep{bilicki_2021} contains $\sim$1 million galaxies and is ideally suited to act as our lens sample. The redshifts ($z_{\rm ANNz2}$) of these galaxies were measured with the \texttt{ANNz2} (artificial neural network) machine learning method \citep{sadeh_2016}. The measurement of these photo-$z$ require a high-completeness spectroscopic training set, for which spectra from the Galaxy And Mass Assembly \citep[GAMA, ][]{driver_2011} survey were used.

%--------------------------------------------------------------------

\section{MICE Tests}
\label{sec:4}

\begin{figure*}
    \centering
    \includegraphics[width=0.96\textwidth]{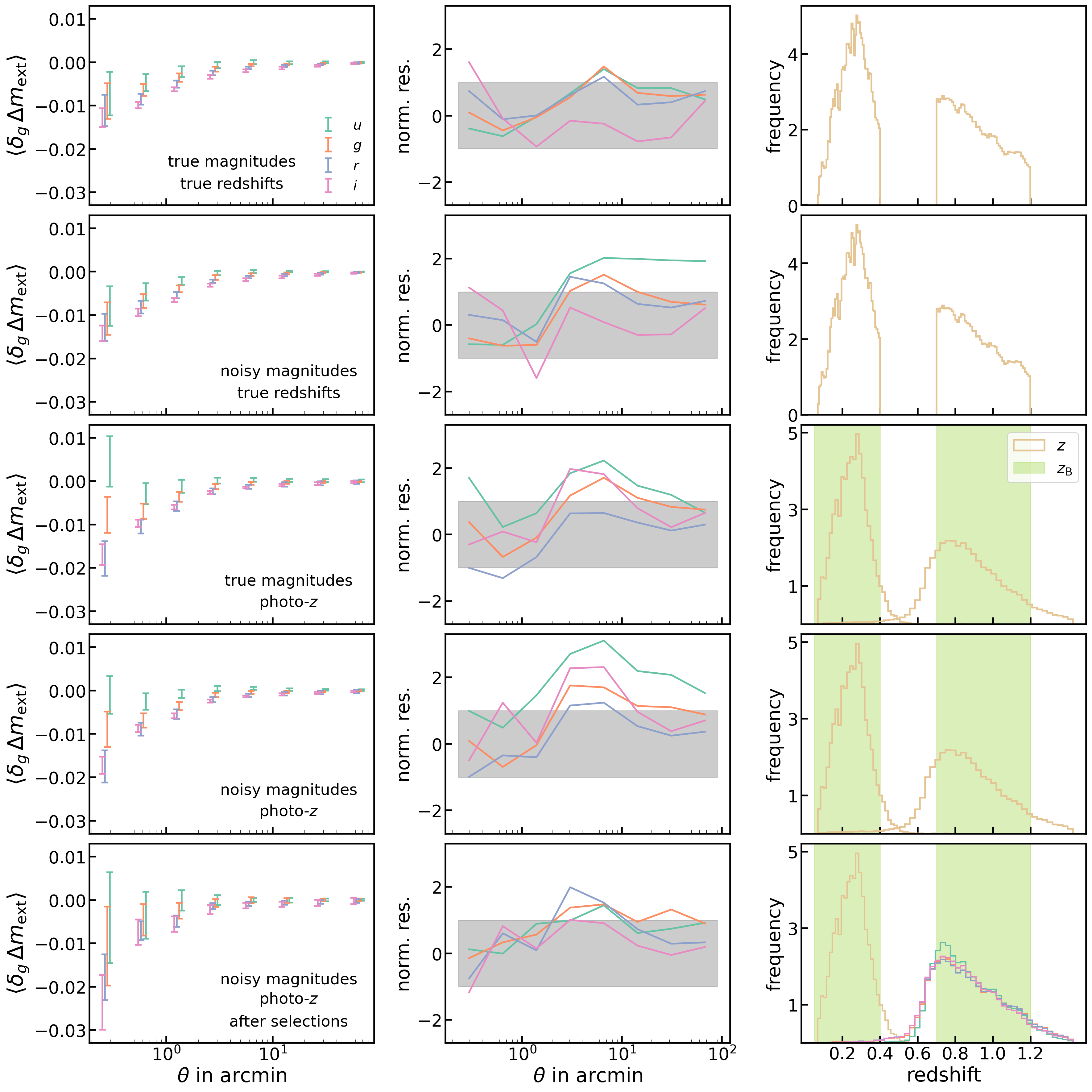}
    \caption{Recovery of magnitude shift signals under increasing observational realism. Each row shows results from a different simulation setup, ranging from ideal conditions (top) to realistic observational effects (bottom). \textit{Left}: Measured magnitude shift signals in KiDS $ugri$ bands. \textit{Middle}: Normalised residuals (measurement minus truth, divided by uncertainty). \textit{Right}: Redshift distributions of lenses (red) and sources (blue). From top to bottom:
(i) Noiseless magnitudes and true redshifts,
(ii) Noisy magnitudes, true redshifts,
(iii) Noiseless magnitudes, photo-$z$,
(iv) Noisy magnitudes and photo-$z$,
(v) Same as (iv), but after applying brightness and template-based source cuts.
Accurate recovery is achieved in the ideal case; signal quality degrades with noise and photo-$z$, especially in the $u$-band. Mitigation strategies restore the signal fidelity in the final row.}
    \label{fig:mice_first_tests}
\end{figure*}

For our cross-correlation measurements, we select galaxies from a contiguous area of $\sim\!\!1000$ deg$^2$ from the simulation octant. Magnitude shift measurements are conducted in eight logarithmic bins between $0.1$ and $100$ arcminutes, while GGL measurements are performed in twelve logarithmic bins between $1$ and $120$ arcminutes. The difference in scales used for the magnitude shift and GGL measurements is driven by the limited resolution of the lensing map in the MICE2 simulation, which restricts our ability to measure the GGL signal on smaller scales (see Sect.~\ref{sec:3} for details). Additionally, the difference in the number of radial bins used for our magnitude shift and GGL measurements is due to the lower signal-to-noise ratio (S/N) of the magnitude shift signal, which therefore requires more pairs per radial bin to achieve a S/N equivalent that measured in GGL. 

We measure both magnitude shift and GGL in ten on-sky patches\footnote{The choice to analyse the signals in patches follows the approach used in MICE2 lensing analyses \citep{fosalba_2015b}, where patching was also applied. We tested the impact of patching on the MICE2 simulation and found no significant difference on scales smaller than 10 arcmin. Additionally, our analysis of KiDS-DR4 data shows that this patching-related difference does not appear in the observational data within the scale range of our analysis.} (each $\sim\!\!100$ deg$^2$) using \texttt{TreeCorr} \citep{treecorr}. For all correlations we apply a random subtraction, designed to suppress constant systematic signals introduced by, for example, survey geometry and observational systematic effects \citep{singh_2017}. Uncertainties are estimated using jackknife resampling, where we recompute our signals after the removal of a single square-degree from the total 1000 deg$^2$ area. These jackknife realisations are then used to construct a covariance matrix for our measurements, which we utilise in our modelling.

\subsection{Systematic effects}
\begin{figure*}
    \centering
    \includegraphics[width=0.9\textwidth]{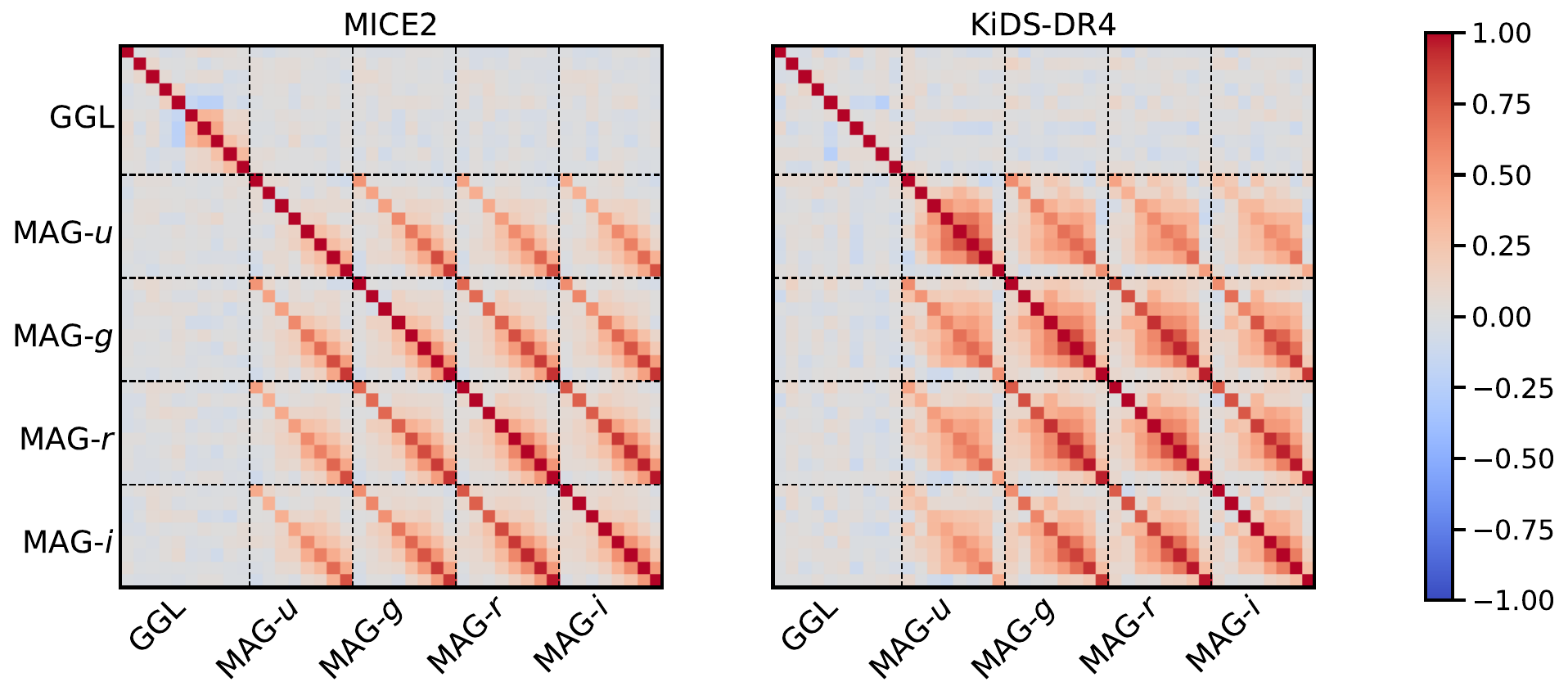}
    \caption{Correlation matrix of the data points used in the MCMC analysis showing the measurements of GGL and magnitude shift signals in the $ugri$ bands, in order. The GGL covariance is largely diagonal and shows little cross covariance with the magnitude shift measurements. Magnitude shift, on the other hand, demonstrates significant cross covariance between the bands, demonstrating the importance of fully modelling the covariance for reddening measurements.}
    \label{fig:corr}
\end{figure*}
We begin our tests by first defining a set of lens and source samples in our simulations, assuming perfect redshift and photometric information. We define our lenses as residing in the redshift range $0.05 < z < 0.4$, and our source sample to reside in the redshift range $0.7 < z < 1.2$. In this way, we ensure that the overlap between the samples in redshift-space is minimised, while retaining a statistically sufficient number of galaxies to preserve the S/N of our measurements. We apply an $i$-band magnitude cut of $m_{\rm evo}^i < 23.5$ to the source galaxy sample, as the MICE2 catalogue is complete down to $i<24$ (as it was designed to mimic a DES-like survey). 
For the lenses, we apply a bright magnitude cut of $m_{\rm evo}^r < 20$, to reproduce a selection similar to the KiDS bright sample. Additionally, we apply a stellar mass requirement to the lens sample of $\log(M_*/M_\odot) > 10.2$, which ensures that our sample of lenses will also produce a strong magnitude shift signal, such that we can clearly observe the effects of reddening. A stronger magnification signal increases the contrast between the intrinsic and observed brightness of background sources, thereby enhancing the detectability of dust-induced reddening. Since reddening is measured by comparing the colours of magnified background galaxies, a clear magnitude shift signature is essential to disentangle the dust extinction effect from statistical noise and other systematics. 

We measure the cross-correlation between the foreground galaxy density field and the residual brightness of sources in the four optical bands of KiDS: $ugri$. Starting from this optimal scenario, we incrementally incorporate observational effects, namely, photometric noise and photometric redshift uncertainties, to assess their impact. The results are summarised in Fig.~\ref{fig:mice_first_tests}. The figure is structured as a $5\times3$ grid, where each row corresponds to a different simulation setup, with increasing observational realism. The left column shows the measured magnitude shift signals, the middle column shows the residuals, i.e. the difference between measured and true signals divided by the uncertainties, and the right column shows the redshift distributions of lenses and sources. To compute the residuals, we simulated the expected magnification signals for each band using the \texttt{Colossus} package, combined with the same extinction profile used in our simulations. Together, these yield the total expected magnitude shift induced on the source galaxies. The magnification signal calculation accounts for the mean halo mass and the median redshift of the lens sample, as well as the critical surface density calculated from the lens and source redshift distributions.

In the first row, we consider the optimal case using noiseless magnitudes and true redshifts. We recover the expected magnitude shift signals in all bands, with residuals mostly within $\pm1\sigma$, confirming that both magnification and reddening effects are captured accurately. The band-dependent differences in signal amplitude, caused by dust extinction, are also recovered. In the second row, we introduce photometric noise to the galaxy magnitudes, while maintaining the true redshifts. Our photometric measurements are realistically modelled in the manner described by \citet{jlvdb_2020}, incorporating consideration of depth, source size, and the effective PSF of imaging made in our set of filters.  Although the overall signal remains visible, the residuals increase, especially in the $u$-band, which is the shallowest and thus most susceptible to flux uncertainties. This highlights the need for applying magnitude cuts to exclude the faintest, and thus the noisiest sources. 

The third row isolates the impact of photometric redshifts, using noiseless magnitudes but replacing the true redshifts with those estimated from multi-band spectral energy distribution (SED) fits to the noisy galaxy fluxes, employing the \texttt{BPZ} code \citep{benitez}, as also described in \citet{jlvdb_2020}. We define lens and source samples using photometric redshifts, with the selected redshift boundaries indicated by green transparent regions in the plots, while the corresponding true redshift distributions are shown in brown. In this case, the signal quality degrades slightly due to the increased uncertainty in the lens–source separation introduced by photometric redshift errors. However, the normalised residuals remain below 2 across all scales, indicating that the redshift boundaries are sufficient to suppress significant contamination from physically associated lens–source pairs.

The fourth row combines both observational magnitudes and photometric redshifts, reflecting a fully realistic measurement scenario without mitigation strategies. The degradation becomes more pronounced, underlining the importance of mitigating both noise and redshift contamination. 

Finally, in the bottom row, we apply distinct magnitude cuts to the source sample.
Photometric noise perturbs the observed galaxy magnitudes and therefore affects our brightness shift measurements. This bias is particularly relevant for faint galaxies, where flux uncertainties are larger. As such, the mitigation strategy for this possible source of bias is relatively straightforward: we can mitigate this effect by applying a brightness cut, thereby limiting our source samples to contain intrinsically less noisy observed fluxes. One must then simply trade off the benefit of more precise photometry with the detriment of smaller source sample sizes. We therefore apply signal-specific brightness cuts to the source galaxies in each filter, excluding the faintest sources to reduce the impact of photometric noise. These cuts allow for a more reliable estimation of the $C_\mathrm{s}$ coefficient for each band. As shown in Appendix \ref{app:intrinsic}, the specific choice of cut does not affect the rest of our analysis, since all downstream quantities are computed consistently within each selection.
These results demonstrate the robustness of our measurement pipeline and highlight the observational challenges that must be addressed in real data.

\begin{table}
    \centering
    \caption{Parameters and their priors used in the MCMC analysis.}
    \begin{tabular}{c c}
        parameter & prior \\ 
        \hline \vspace{1.6mm}
        log$\left(M_{\rm halo}^{\rm DM}/M_\odot\right)$ & [10, 15] \\ \vspace{1.7mm}
        log$\left(M_{\rm halo}^{\rm dust}/M_\odot\right)$ & [6, 10] \\
        $\delta_1$ & [1,100]  \\
        $s$ &  [0.1,4]\\
        \hline
    \end{tabular}
    \label{tab:priors}
\end{table}

\begin{table*}
\label{tab:mcmc_m}
\caption{Parameter constraints from the MCMC analysis based on MICE2 data, including results from GGL only, magnitude shift (in the $ugri$ bands) only, and the joint fit combining both.}
\centering
\begin{tabular}{cccccccc}
\hline\hline 
\noalign{\vskip 0.2em}
Probe & log($M_{\rm halo}^{\rm DM}$) & log($M_{\rm halo}^{\rm dust}$) & $\delta_1$&$s$ &$\chi^2$ &$\nu$&PTE\\ [0.2em]
\hline 
\noalign{\vskip 0.3em}
Simulation & $12.32$ &$7.7$ & & &  &  & \\
GGL-only & $12.34^{+0.05}_{-0.09}$ &$7.05^{+2.12}_{-0.82}$ &$34.01^{+6.98}_{-3.52}$ & $1.47^{+0.05}_{-0.05}$&7.80&6&0.25\\[0.5em]

$\mu$-only&$12.35^{+0.05}_{-0.30}$  &$7.65^{+0.05}_{-0.28}$&  $60.19^{+27.87}_{-17.01}$  & $1.65^{+1.10}_{-0.04}$&22.91 &28& 0.73\\[0.5em]

Joint&$12.33^{+0.04}_{-0.08}$   &$7.70^{+0.03}_{-0.05}$&$43.79^{+11.04}_{-4.05}$  & $1.54^{+0.05}_{-0.02}$&30.55&38&0.80 \\
\noalign{\vskip 0.3em}
\hline
\end{tabular}
\tablefoot{For each analysis, the table lists the best-fitting parameter values along with the 68\% credible intervals, as well as the corresponding chi-square values ($\chi^2$), number of degrees of freedom ($\nu$), and PTE values assessing the goodness of fit. We note that the dust mass in GGL-only case is prior dominated. We also present the mean halo mass of the lens galaxy sample and the input dust mass as reference.}
\end{table*}

\begin{figure}
    \centering
    \includegraphics[width=\columnwidth]{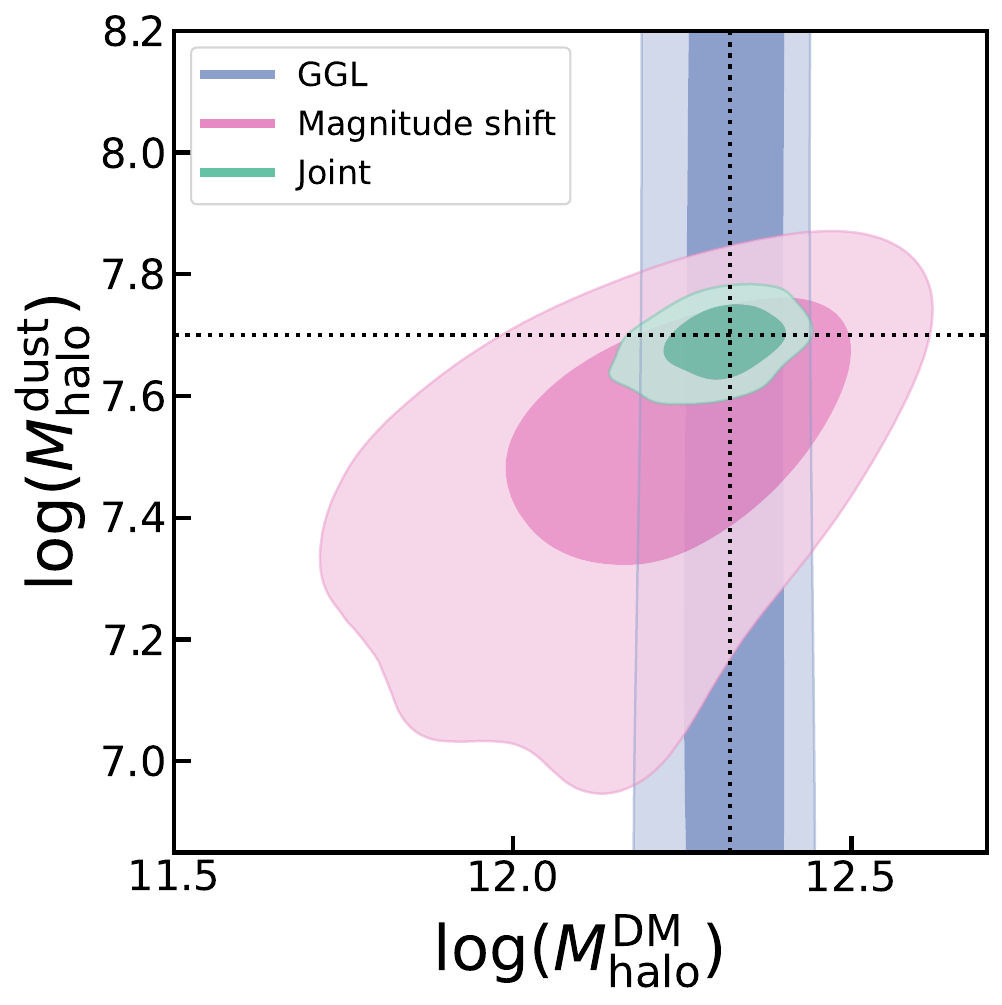}
    \caption{Confidence regions from the MCMC analysis of noisy MICE2 galaxies in the log($M_{\rm halo}^{\rm DM}$)-log($M_{\rm halo}^{\rm dust}$) parameter space. The contours represent the 1$\sigma$ and 2$\sigma$ confidence intervals for the posterior distributions of the model parameters. Dotted lines show the true mean halo mass of our lens sample and the input dust mass. While GGL provides strong constraints on halo mass but no constraints on dust mass, magnitude shift constrains both parameters, though with lower precision. By combining both measurements in a joint analysis, we recover both parameters within the 1$\sigma$ interval. This result validates our measurement pipeline using the MICE2 simulation, which will be applied to the KiDS DR4 dataset.}
    \label{fig:mice_mcmc_contours}
\end{figure}

\begin{figure*}
    \centering
    \includegraphics[width=\textwidth]{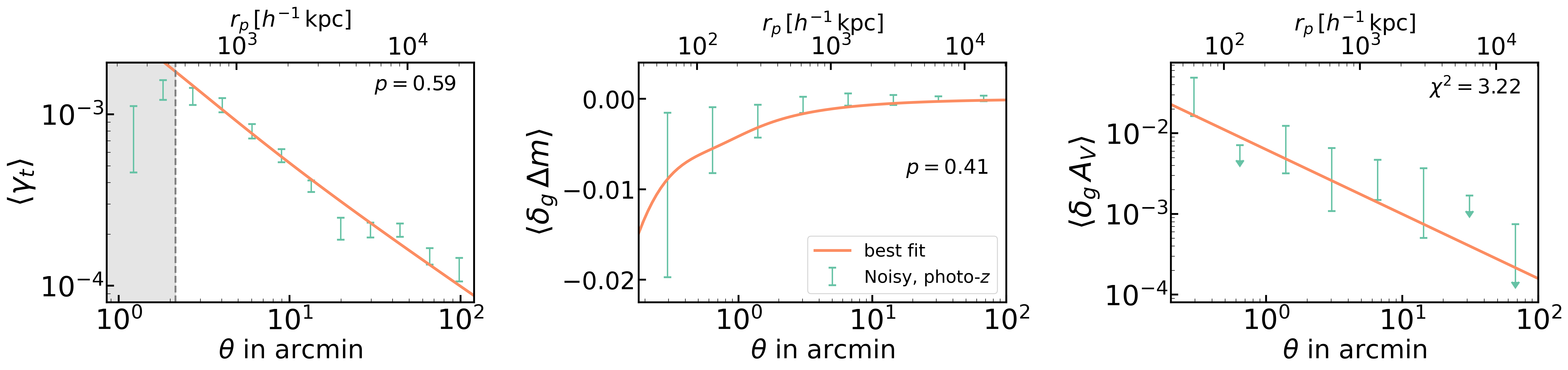}
    \caption{Validating our measurement pipeline with the MICE2 mock galaxy catalogue; from left to right; GGL, magnitude shift ($g$-band), and reddening ($g-i$ colour) measurements. Purple data points show the measurement on noiseless data (noiseless magnitudes and true redshifts), green depicts the measurement on noisy data (noisy magnitudes and photometric redshifts) after selection criteria applied, and the orange lines are the best-fits (maximum posterior) resulting from MCMC analysis. Noisy data for the GGL measurement refers to the use of galaxies with shape noise and photometric redshifts. Shaded grey area depicts the region, which is excluded from the MCMC analysis due to poor resolution. Our measurements are in good agreement with best fits obtained from the MCMC analysis.}
    \label{fig:mice_ppds}
\end{figure*}

\subsection{Estimation of halo mass and dust mass}

Once we established our selection criteria for lens and source samples, we perform our cross-correlation measurements for the most realistic observational case, and run a Markov Chain Monte-Carlo (MCMC) analysis to find the best parameters in our halo and extinction model. In doing so, we are able to test the recovery of input halo and and circumgalactic dust masses for our sample of lens galaxies. 

We implement a standard Gaussian likelihood in our MCMC analysis, with fully specified covariance matrix (including cross-covariance between statistics, where appropriate): 
\begin{equation}
    \mathfrak{L} \propto [\boldsymbol{\xi}-\boldsymbol{\xi}_\mathrm{m}]^\intercal\mathrm{C}^{-1}[\boldsymbol{\xi}-\boldsymbol{\xi}_\mathrm{m}].
\end{equation}
Here, $\boldsymbol{\xi}$ is the data vector of correlation functions measured on galaxies in observational case, $\boldsymbol{\xi}_\mathrm{m}$ is the data vector of correlation functions predicted from the model during the MCMC analysis, and $C$ denotes the covariance matrix, which is estimated with jackknife resampling over $1000$ patches using \texttt{TreeCorr} \citep{treecorr}, with each patch being $1\,\mathrm{deg}^2$. We also applied the so-called Hartlap correction to our covariance matrix, which is needed as the inverse of a sample covariance matrix is biased when estimated from a finite number of simulations \citep{hartlap_cov}. The resulting correlation matrix for MICE2 can be found in Fig.~\ref{fig:corr} on the left hand side. The covariance matrix of GGL is largely diagonal, primarily dominated by shape noise, and has little to no cross-covariance with magnitude shift. Conversely we observe significant cross-covariance between our individual magnitude shift measurements in each band. These arise from several contributing factors. First, the variances in magnitude errors in different bands are correlated, due to the underlying SED of the sources. Second, the sample variance component to the cross-covariance is significant, due to the achromatic nature of the magnification signal. Finally, magnitude shift signals measured in different angular bins can trace the same underlying lensing potential, leading to cross-correlations between angular scales and across filters. 
The significant cross-covariance between the magnitude shift measurements is of particular importance for our reddening estimates, as the reddening is directly computed from differences in magnitude shift models.

We modelled the matter distribution around lens galaxies using a two-component halo model to construct the likelihood function used in our MCMC analysis. The 1-halo term is described by an NFW profile, while the 2-halo term accounts for the clustering of haloes in large-scale structures, through an infalling component following the formulation of \citet[][Eq.~\ref{eq:rho}]{diemer_2023}. This halo model is used to predict the excess surface mass density profiles ($\Delta\Sigma$), which are directly compared to our GGL measurements. The same halo model is also used to compute the surface mass density profiles ($\Sigma$) when modelling the magnification signal. 

To account for extinction effects in the magnitude shift measurements, we incorporate a dust component into the model. Given a dust mass, the extinction profile in the $V$-band is computed using Eq.~\eqref{eq:dust_mass}, assuming a fixed slope of $-0.8$ for the extinction profile in $V-$band, which is derived from the observational results of \cite{menard_2010}. Because the slope is fixed, the dust mass $M_{\rm halo}^{\rm dust}$ directly determines the amplitude of the extinction profile, which allows us to treat $M_{\rm halo}^{\rm dust}$ as the free parameter in our MCMC analysis and convert it into an extinction amplitude. We also note that we adopt the absorption optical depth per unit dust mass as $K_{\rm ext} = 3.217\, \mathrm{pc}^2 M_\odot^{-1}$, consistent with SMC-type dust as described in \citet{w_draine_2001} and also used in \citet{menard_2010}.
The modelled $V$-band extinction is converted to other photometric bands using the extinction law of \citet{fitzp_1999}, in order to allow comparison with observed extinction signals, assuming a Milky Way-type extinction curve with $R_V = 3.1$. Finally, these rest-frame values are shifted to the observer frame using the median redshift of the lens sample.

In the halo model, the halo mass, the dust mass, and the parameters of $\delta_1$ and $s$ are left free. Our data vectors include GGL signals and magnitude shift measurements made in our four optical bands $ugri$. Since the reddening information is already incorporated implicitly in the various magnitude shift measurements, we do not include it separately in our MCMC analysis (indeed, including the extinction explicitly would result in a non-invertible covariance matrix). 
In Table \ref{tab:priors}, we present the parameters of our fitting model used in the likelihood analysis, and their priors. For the priors of parameters $\delta_1$ and $s$, we used the lower and upper limits of the profile provided by \citet{diemer_2023}.

We use the ensemble sampler \textit{emcee} \citep{emcee} to explore the posterior distributions of our model parameters. The MCMC was run with 50 walkers and 1000 steps per walker, while we remove the first 150 steps of each walker as burn-in (and validate the burn-in removal visually). To assess convergence, we performed visual inspection of the trace plots to ensure proper mixing of the walkers and examined the corner plot to confirm that the posterior distributions are unimodal and well-sampled. Example posteriors are provided in Appendix \ref{app:full_mcmc_constraints}. For parameter inference, we identify the best-fit point as the maximum posterior sample in the converged chain. 

We present the $1\sigma$ and $2\sigma$ confidence intervals for the posterior distributions of the halo mass and the dust mass in Fig.~\ref{fig:mice_mcmc_contours}, where the true mean halo mass and the simulated dust mass are shown with dotted lines. The mean halo mass is calculated as the mean of the individual galaxy halo masses weighted by the halo mass function, as the lensing signal is not linear with halo mass \citep{dvornik_2017}:
\begin{equation}
   \langle M_{\rm halo} \rangle = \frac{\int M_{\rm halo} \, n(M_{\rm halo}, z) \, \rm{d}M_{\rm halo}}{\int n(M_{\rm halo}, z) \, \rm{d}M_{\rm halo}}.
\end{equation}
The true dust mass is calculated with the simulated extinction profile as given in Eq.~\eqref{eq:dust_mass}.

In Table~\ref{tab:mcmc_m}, we present the parameter constraints from the MCMC analysis of MICE2 data with three different probes: GGL only, magnitude shift only, and the joint fit. The recovered values are fully consistent with the input values within $1\sigma$, confirming the accuracy of our model. As expected, GGL alone constrains the halo mass well but is insensitive to the dust mass. In contrast, magnitude shift provides constraints on both parameters, but the degeneracy between halo mass and dust mass results in poorer constraints on the halo mass compared to GGL. The joint analysis of GGL and magnitude shift, however, breaks this degeneracy by leveraging halo-mass information from the dust-independent GGL signal, thereby enabling more precise and reliable estimates of both halo and dust mass. 

Examples of our measured signals can be found in Fig.~\ref{fig:mice_ppds}, where we present the $g$-band signal for magnitude shift and the $V-$band extinction profile calculated from $g-i$ colour excess (We refer the reader Fig.~\ref{fig:mice_magn_ugri} for the results of other bands.). Green data points depict the measurements made using our most realistic simulations of the observational data, while the orange line shows the best fit resulting from MCMC analysis. The grey area indicates scales excluded from our analysis. In the case of extinction, these data are not explicitly used in the calculation of our best-fit models via MCMC, as the extinction data vector is constructed from combinations of magnitude shift signals, and thus is fully degenerate with magnitude shift. Including such signals in our optimisation (and thus also in our covariance) would result in an non-invertible covariance, and so are not used directly. Nonetheless, we show the effective data-vector for extinction in the figure (and elsewhere), as a direct visualisation of the estimated extinction returned by our modelling of GGL and multi-band magnitude shift. The uncertainties on the extinction signals in the figure are constructed by a straightforward error propagation of the diagonal of the covariance on the individual magnitude shift signals that have been used; while this is technically incorrect, these uncertainties are only used for visualisation purposes and the computation of an approximate goodness-of-fit (below).

Our measurements show good agreement with the final model fits in all bins. Our data is well described by our fitted models, as evidenced by a passing goodness-of-fit per statistic and overall (as reported by the null hypothesis probability-to-exceed, PTE). In Fig.~\ref{fig:mice_ppds}, we annotate the GGL and magnitude shift panels with the PTE estimated for the data-vector contained within that panel, computed using the best fit model from the joint fit to all observables. We find that all statistics (shown and not shown) are well fit by the joint model (with $p>0.01$). 
For the reddening signal, we do not report the PTE as we have no appropriate estimate of the number of degrees of freedom from this data vector (given the covariance between the magnitude shift signals). As such, we report only a simple $\chi^2$ estimated using propagation of the uncertainties on each of our magnitude shift signals (assuming independence), yielding $\chi^2 = 3.22$.

\begin{figure}
    \centering
    \includegraphics[width=\columnwidth]{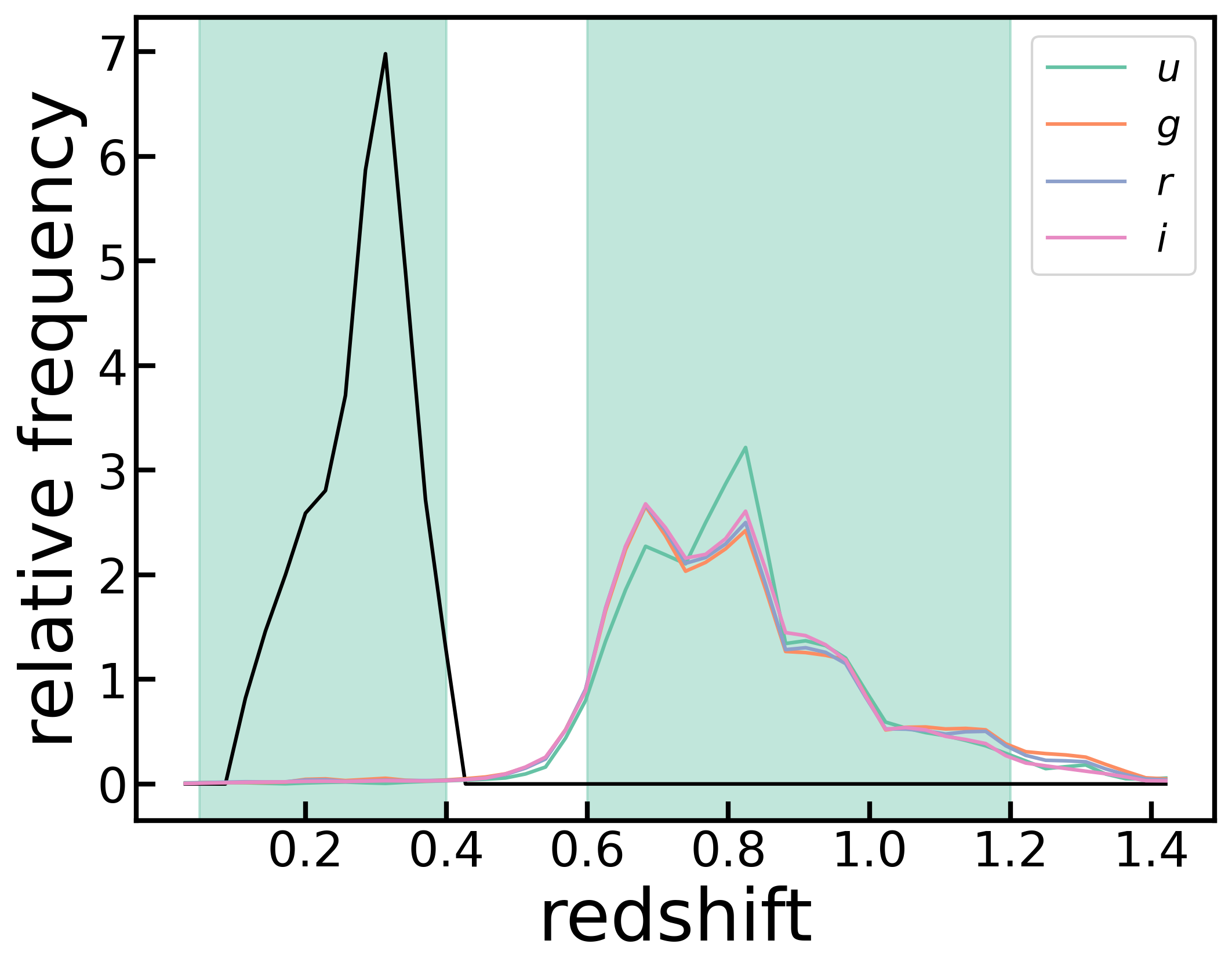}
    \caption{We show source redshift distributions corresponding to different magnitude cuts in each band. As discussed in the main body of the paper, using band-specific $N(z)$ estimates yields negligible differences in the resulting measurements. For consistency, a single $N(z)$ (the one with $g-$band cut) is used in the main analysis.}
    \label{fig:kids_zdist}
\end{figure}

\begin{figure}
    \centering
    \includegraphics[width=\columnwidth]{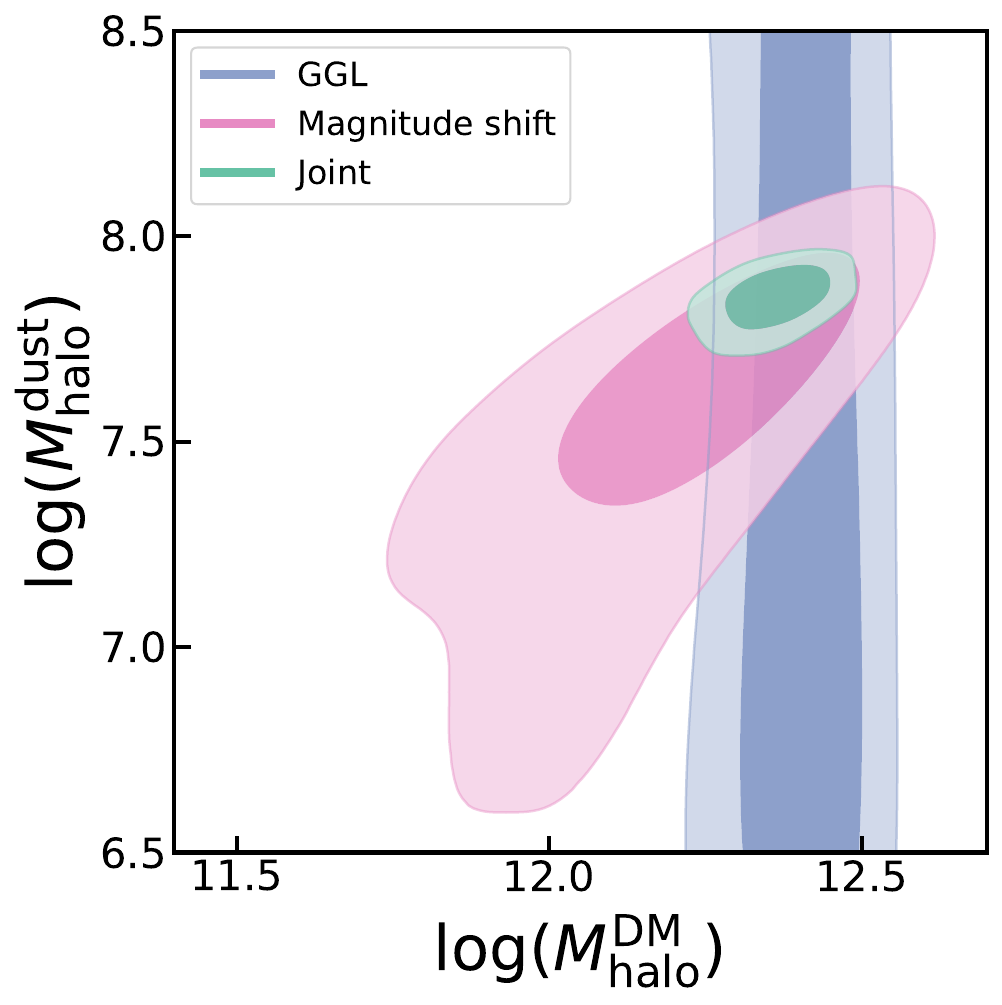}
    \caption{Confidence regions from the MCMC analysis of KiDS-DR4 galaxies in the log($M_{\rm halo}^{\rm DM}$)-log($M_{\rm halo}^{\rm dust}$) parameter space. Compared to the MICE2 analysis, magnitude shift only measurement yields broader confidence contours. As expected, the joint analysis with GGL significantly tightens the constraints on both parameters, demonstrating the power of combining both probes.}
    \label{fig:kids_mcmc_contours}
\end{figure}

\begin{table*}
\caption{\label{tab:mcmc_k}Parameter constraints from the MCMC analysis based on KiDS-DR4 data, including results from GGL only, magnitude shift (in the $ugri$ bands) only, and the joint fit combining both.}
\centering
\begin{tabular}{cccccccc}
\hline\hline 
\noalign{\vskip 0.2em}
Probe & log($M_{\rm halo}^{\rm DM}$) & log($M_{\rm halo}^{\rm dust}$) & $\delta_1$&$s$ &$\chi^2$ &$\nu$&PTE\\ [0.2em]
\hline 
\noalign{\vskip 0.3em}

GGL-only & $12.35^{+0.13}_{-0.05}$ &$7.84^{+1.11}_{-2.08}$ &$99.36^{+0.20}_{-26.44}$ & $1.79^{+0.04}_{-0.09}$&6.16&7&0.52 \\[0.5em]

$\mu$-only&$12.30^{+0.08}_{-0.24}$  &$7.65^{+0.21}_{-0.27}$&  $87.68^{+6.06}_{-28.65}$  & $1.82^{+0.16}_{-0.08}$&20.16 &28& 0.86\\[0.5em]

Joint&$12.33^{+0.08}_{-0.04}$   &$7.86^{+0.03}_{-0.04}$&$98.56^{+1.42}_{-19.44}$  & $1.80^{+0.01}_{-0.07}$& 30.77 &39&0.82\\
\noalign{\vskip 0.3em}
\hline
\end{tabular}
\tablefoot{For each analysis, the table lists the best-fitting parameter values along with the 68\% credible intervals, as well as the corresponding chi-square values ($\chi^2$), number of degrees of freedom ($\nu$), and PTE values assessing the goodness of fit. We note that the dust mass is prior dominated, when we use only the GGL signal.}
\end{table*}

\begin{figure*}
    \centering
    \includegraphics[width=\textwidth]{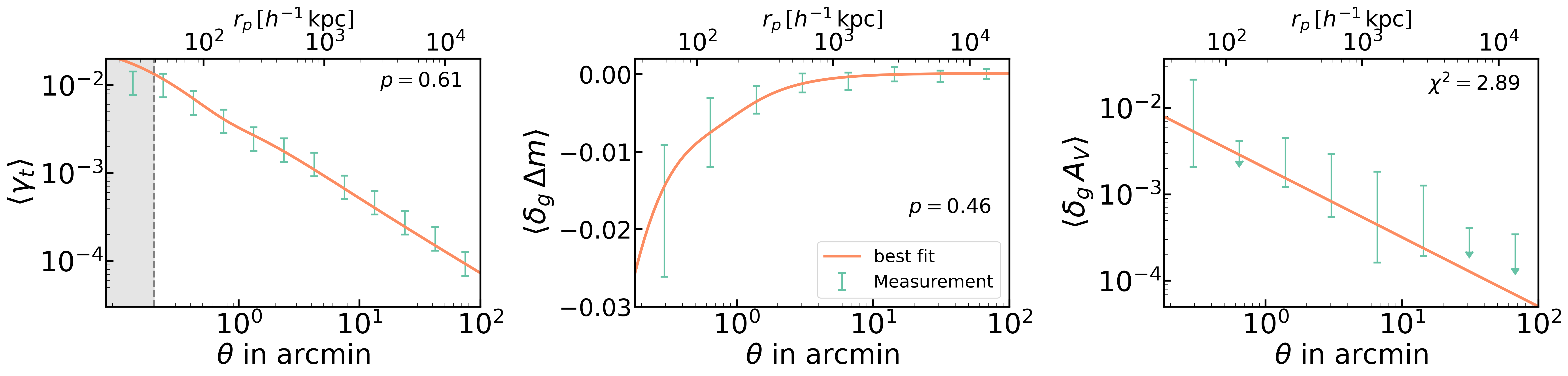}
    \caption{Similar to Fig.~\ref{fig:mice_ppds}, we present the GGL measurement, magnitude shift in the $g-$band and the extinction profile in $V-$band calculated from the $g - K_{\rm s}$ colour. Our measured signals match closely with best fits from the MCMC analysis. Overall, the results indicate that our model provides a good fit to the measured signals.}
    \label{fig:kids_ppds}
\end{figure*}

\section{KiDS-DR4 Results}
\label{sec:5}

Our analysis of the MICE2 simulations provides confidence in the analysis framework that we have constructed, from estimation of data vectors and covariances through to the estimation of physical parameters for the sample of lens galaxies under analysis. Having validated our measurement pipeline, we now move to an analysis of real-world data from KiDS-DR4. 

As described in Sect.~\ref{sec:3}, we selected source galaxies from the KiDS-1000 weak lensing gold catalogue, and lens galaxies from the KiDS-bright sample of galaxies.
The lens sample consists of galaxies with $0.05 < z_{\rm ANNz2} < 0.4$, $m_r < 20$, and $\log(M_*/M_\odot) > 10.2$. Source galaxies are selected in the range $0.7 < z_B < 1.2$, with $m_x < m_{\rm faint}$, where $x$ indicates the photometric band and $m_{\rm faint}$ the corresponding band-specific magnitude cut. We estimated the true redshift distributions of the source samples $N(z)$ using self-organising maps (SOMs) and found that the resulting variations in $\Sigma_{\rm crit}$ between samples remain below $0.5\%$. Consequently, we adopt the $N(z)$ of the $g$-band selected sample for the full analysis, and verified that repeating the procedure with the other $N(z)$ estimates yields only negligible differences. These correspond to systematic biases of at most $10\%$ of the magnitude shift uncertainties in all bands, as the induced variations in $\Sigma_{\rm crit}$ remain below $0.5\%$.
The redshift distributions of lens and source galaxies are shown in Fig.~\ref{fig:kids_zdist}.

Utilising the analysis pipeline from MICE2, we perform an equivalent measurement of GGL, and magnitude shift in four optical bands ($ugri$). We measure the mean halo mass and the dust mass of our lens sample via a joint analysis of GGL and magnitude shift with an MCMC analysis, using the same priors and analysis settings (i.e. MCMC settings and post-processing of posteriors) as in our analysis of MICE2. The covariance matrix is similarly constructed in the same manner for KiDS as in MICE2, via jackknife resampling over 1000 patches, with each being 1 deg$^2$. The correlation matrix for the KiDS data is qualitatively similar to that measured on MICE2 and shown in Fig.~\ref{fig:corr}, with considerable off-diagonal terms in magnitude shift. The overall covariance in KiDS is larger than in our simulated case, which we attribute to a larger effective variance in our overall data coming from the observational complexities that are not represented in our simulations (in particular those related to the analysis of images, such as image reduction, source extraction, source definition, and photometric measurement). While this discrepancy highlights that our simulations underestimate the true observational noise, it does not invalidate our results. The key reason is that our parameter inference is based on the data covariance, not the simulated one. The increased covariance in the real data naturally leads to broader posteriors, which we properly account for in our likelihood analysis. Therefore, although the reduced precision implies a lower S/N compared to the ideal case, our conclusions remain robust because the uncertainties are correctly propagated.

We present the posterior constraints on halo mass and dust mass from our three analyses of KiDS data in Fig.~\ref{fig:kids_mcmc_contours}, and tabulated summaries of the posteriors in Table~\ref{tab:mcmc_k}. We find that the parameter constraints estimated from optimisations using GGL only, magnitude shift only, and joint fits to both sets of data are fully consistent with each other. Furthermore, in Table~\ref{tab:mcmc_k} we present the overall PTE values for each analysis setup, finding that all data vectors are well fit by their respective models. As in the case of MICE2, our joint analysis of GGL and magnitude shift  significantly tightens the constraints compared to magnitude shift alone. 
We constrain marginal estimates for halo mass and dust mass of the KiDS-bright lens sample to $\log\left(M_{\rm halo}^{\rm dust}/M_\odot\right)=7.86^{+0.03}_{-0.04}$ and $\log\left(M_{\rm halo}^{\rm DM}/M_\odot\right)=12.33^{+0.08}_{-0.04}$.

As in our analysis of MICE2, we present the measured valued and the best fits of our joint fits of GGL and magnitude shift in Fig.~\ref{fig:kids_ppds}, and again report the PTE values derived from the joint model fit for GGL and magnitude shift. We note that, instead of the $g - i$ reddening signal, we show the extinction profile in $V-$band calculated from $g - K_{\rm s}$ colour excess. Overall, all goodness-of-fit values all indicate that our model provides a statistically robust description of the data.

\subsection{Comparisons to literature constraints}

\begin{figure}
    \centering
    \includegraphics[width=\columnwidth]{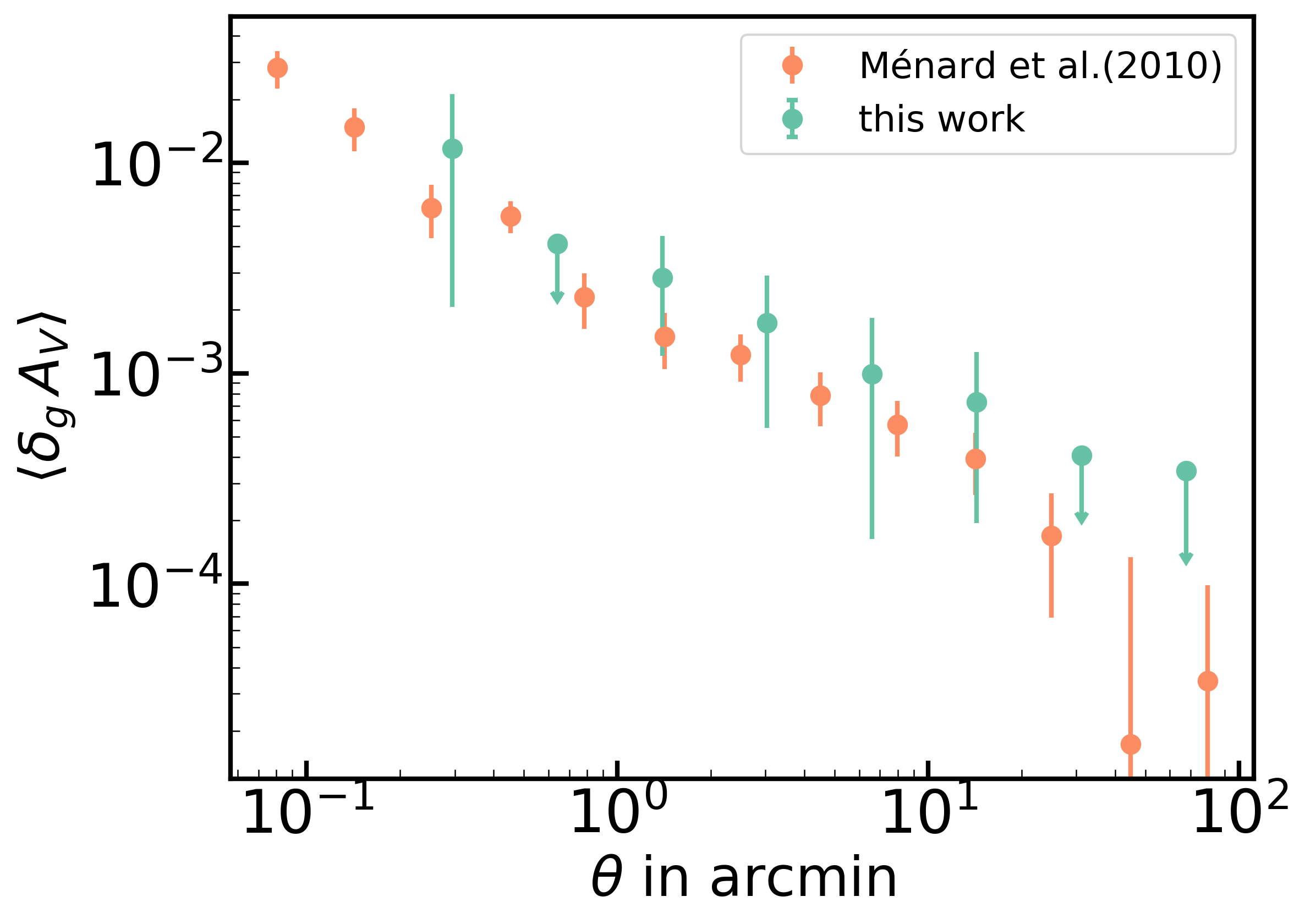}
    \caption{Comparison of our measured extinction profile $A_V$ with the results of \citet{menard_2010}. We calculated the extinction profile from the $g - K_\mathrm{s}$ colour excess measurement, as it has the highest S/N among other colour combinations.}
    \label{fig:5_4}
\end{figure}

Our values for both halo mass and dust mass are consistent with previous observational and theoretical studies. Since our lens sample has a comparable mean redshift ($z\sim0.3$) to that of \citet{menard_2010} ($z\sim0.36$), we can directly compare our extinction profiles. We present the extinction profile in $V-$band calculated from the measured $g-K_{\rm s}$ colour excess signals, alongside the extinction profile measured by \citet{menard_2010} in Fig.~\ref{fig:5_4}. We see that our profiles closely match with each other within the error bars, which is a result of the comparable dust-to-stellar mass ratio: We find a ratio of 
\( M_{\rm dust}/M_* \approx 1.9 \times 10^{-3} \) in our sample, which is highly consistent 
with the value of \( 2.0 \times 10^{-3} \) reported by \citet{menard_2010}. We note that we present here the signal with the highest S/N among the other colour combinations, which can be found in Fig.~\ref{fig:app_7}.
The S/N of $g - i$ colour excess signal of \citet{menard_2010} is about 2.6 times higher than ours, which can be attributed to the broader colour distributions of our source galaxies. More details and supporting plots of the magnitude and colour distributions are presented in Appendix \ref{app:colordist}.

Our measurements also align well with the value reported by \citet{peek_2015}, who found $6 \pm 2 \times 10^7\,M_\odot$ around SDSS galaxies at $z \sim 0.05$ using colour excess measurements, implying a dust-to-stellar mass ratio of 
\( 1.0 \times 10^{-3} \). Our work, \citet{menard_2010}, and \citet{peek_2015} all use low-redshift galaxies with similar mean stellar masses and adopt the same SMC-type dust opacity, enabling a direct comparison and yielding consistent dust mass estimates. These comparable ratios suggest that the dust content relative to stellar mass in galaxies remains fairly consistent across different samples and methodologies at similar redshifts.
In addition, theoretical predictions from the SAGE semi-analytical model with a detailed dust prescription \citep{triani_2020} also support the presence of such CGM dust masses at $z \sim 0$, further reinforcing the observational evidence for a diffuse dust component in galactic halos. 

Overall, our estimated physical parameters from KiDS agree well with existing literature on low-redshift samples of galaxies. In future work, we will extend this analysis to subsamples of KiDS, in an effort to explore these properties as a function of stellar mass, star formation rate, and photometric redshift. We will present the results in a follow-up paper. A key challenge in this extended analysis is the disentanglement of redshift-dependent effects in the interpretation of the results, e.g. in constraining the relationship between dust mass and stellar mass across different epochs.

%-----------------------------------------------------------------

\section{Conclusions}
In this study, we aim to quantify the amount and distribution of dust in galactic halos and to measure the mean halo mass and mean dust mass in the halos through a joint analysis of GGL and multi-band magnitude shift signals. To this end, we used the KiDS-DR4 data, which provides accurate redshift estimates, reliable shape measurements, and high-precision galaxy photometry \citep{kuijken_2019}. Following the methodology of \citet{menard_2010}, we exploited both GGL and magnitude shift signals, with the addition of multi-band data allowing us to disentangle extinction effects. In contrast to \cite{menard_2010}, we used galaxies as sources instead of QSOs. Using galaxies as source sample introduces specific challenges, which is why we first used the MICE2 mock galaxy catalogue to investigate the issues and to establish selection criteria for lens and source galaxies.

As the MICE2 catalogue does not account for extinction effects caused by circumgalactic dust, we simulated this effect based on the findings of \cite{menard_2010}, and modified the galaxy magnitudes accordingly. To ensure consistency with their study, we restricted our lens sample to galaxies with stellar masses $\mathrm{log}(M_*/M_\odot) > 10.2$, so that the mean stellar mass would approximately match that of their sample.

In our analysis with the MICE2 galaxies, we identified photometric redshift uncertainties as a primary challenge. By defining lens and source samples with a substantial redshift separation ($\Delta z \sim 0.3$), we largely eliminate redshift overlap. This separation suppresses contamination from physically associated lens–source pairs, which could otherwise introduce intrinsic magnitude correlations. To further mitigate observational effects, we applied magnitude cuts to the source sample in each band, excluding the faintest galaxies where photometric noise dominates. This strategy allows us to maintain a reliable measurement of magnitude shifts. Our tests confirm that, with these selections, the residual contamination is negligible, demonstrating that our pipeline is robust and capable of capturing the combined effects of magnification and dust reddening in realistic survey conditions.

Our MCMC analysis using the MICE2 data demonstrates that we can recover the mean halo mass and input circumgalactic dust mass with high accuracy and precision through a joint analysis of GGL and magnitude shift signals across four optical bands. The GGL signal plays a crucial role in this analysis, as it is less sensitive to photometric redshift uncertainties and the resulting redshift overlap between lens and source galaxies. This makes GGL a more robust probe of the halo mass, allowing for tighter constraints compared to magnitude shift. In turn, this helps disentangle the effects of dust extinction from magnitude shift signals, enabling a more reliable extraction of the circumgalactic dust component.

After applying our measurement pipeline to the KiDS-DR4 data, we obtained a dust mass of $7.24\times10^7\,M_\odot$ and a dust-to-mass ratio of $3.3\times10^{-5}$ for galaxies at $\langle z\rangle\sim0.3$ with stellar mass $\mathrm{log}(M_*/M_\odot)>10.2$. Our results are in good agreement with \citet{menard_2010} and \citet{peek_2015}, and consistent with the predictions of the SAGE model from \citet{triani_2020}. Remarkably, this agreement is achieved despite significant differences in methodology—ranging from galaxy–QSO pair counts to color excess profiles and semi-analytical modeling—highlighting the robustness of the inferred CGM dust content.

In future work, we will investigate the dependence of dust profiles on the halo mass, stellar mass, and star formation rate of galaxies. The results of such analyses could provide valuable insight into questions related to galaxy evolution and cosmology.
However, several challenges must be addressed to fully exploit these measurements. First, the current data do not allow us to robustly constrain the absorption coefficient per unit dust mass, which directly affects the measured dust mass. Moreover, measurements on smaller scales ($<30\,h^{-1}\,\mathrm{kpc}$) could reveal crucial details about the mass and distribution of CGM dust, but these are inaccessible due to the increasing impact of intrinsic correlations. Such intrinsic correlations must be carefully accounted for when using galaxies as background sources in magnitude shift studies. Additionally, we find that the covariance of the magnitude shift signals is highly significant, highlighting the necessity of accurate and robust covariance matrix estimation for reliable inference.

\begin{acknowledgements}
       We would like to thank Andrej Dvornik, Jan Luca van den Busch, Dillon Brout, Brice Ménard, Shiyang Zhang, and Anna Wittje for useful discussions during the project. EG acknowledges the support from the Deutsche Forschungsgemeinschaft (DFG) SFB1491. AHW is supported by the Deutsches Zentrum für Luft- und Raumfahrt (DLR), made possible by the Bundesministerium für Wirtschaft und Klimaschutz, under project 50QE2305, and acknowledge funding from the German Science Foundation DFG, via the Collaborative Research Center SFB1491 "Cosmic Interacting Matters - From Source to Signal". H. Hildebrandt is supported by a DFG Heisenberg grant (Hi 1495/5-1), the DFG Collaborative Research Center SFB1491, an ERC Consolidator Grant (No. 770935), and the DLR project 50QE2305.
       The data used in this work are based on observations made with ESO Telescopes at the La Silla Paranal Observatory under programme IDs 177.A-3016, 177.A-3017, 177.A-3018 and 179.A-2004, and on data products produced by the KiDS Consortium.  The KiDS production team acknowledges support from the DFG, ERC, NOVA and NWO-M grants; Target; the University of Padova, and the University Federico II (Naples). This work was supported by a grant of the German Centre of
  Cosmological Lensing, hosted at Bochum University.

\end{acknowledgements}

\bibliographystyle{aa}
\bibliography{References}

\begin{thebibliography}{41}
\expandafter\ifx\csname natexlab\endcsname\relax\def\natexlab#1{#1}\fi

\bibitem[{Aguirre {et~al.}(2001)Aguirre, Hernquist, Schaye, Weinberg, Katz, \& Gardner}]{ag_2001}
Aguirre, A., Hernquist, L., Schaye, J., {et~al.} 2001, \apj, 560, 599

\bibitem[{{Asgari} {et~al.}(2021){Asgari}, {Lin}, {Joachimi}, {Giblin}, {Heymans}, {Hildebrandt}, {Kannawadi}, {St{\"o}lzner}, {Tr{\"o}ster}, {van den Busch}, {Wright}, {Bilicki}, {Blake}, {de Jong}, {Dvornik}, {Erben}, {Getman}, {Hoekstra}, {K{\"o}hlinger}, {Kuijken}, {Miller}, {Radovich}, {Schneider}, {Shan}, \& {Valentijn}}]{asgari2021}
{Asgari}, M., {Lin}, C.-A., {Joachimi}, B., {et~al.} 2021, \aap, 645, A104

\bibitem[{{Bartelmann} \& {Schneider}(2001)}]{bartelmann_2001}
{Bartelmann}, M. \& {Schneider}, P. 2001, \physrep, 340, 291

\bibitem[{{Ben{\'\i}tez}(2000)}]{benitez}
{Ben{\'\i}tez}, N. 2000, \apj, 536, 571

\bibitem[{{Bianchi} \& {Ferrara}(2005)}]{bianchi_2005}
{Bianchi}, S. \& {Ferrara}, A. 2005, \mnras, 358, 379

\bibitem[{{Bilicki} {et~al.}(2021){Bilicki}, {Dvornik}, {Hoekstra}, {Wright}, {Chisari}, {Vakili}, {Asgari}, {Giblin}, {Heymans}, {Hildebrandt}, {Holwerda}, {Hopkins}, {Johnston}, {Kannawadi}, {Kuijken}, {Nakoneczny}, {Shan}, {Sonnenfeld}, \& {Valentijn}}]{bilicki_2021}
{Bilicki}, M., {Dvornik}, A., {Hoekstra}, H., {et~al.} 2021, \aap, 653, A82

\bibitem[{{Capaccioli} {et~al.}(2005){Capaccioli}, {Mancini}, \& {Sedmak}}]{capaccioli_2005}
{Capaccioli}, M., {Mancini}, D., \& {Sedmak}, G. 2005, The Messenger, 120, 10

\bibitem[{{Carretero} {et~al.}(2015){Carretero}, {Castander}, {Gazta{\~n}aga}, {Crocce}, \& {Fosalba}}]{Carretero_2015}
{Carretero}, J., {Castander}, F.~J., {Gazta{\~n}aga}, E., {Crocce}, M., \& {Fosalba}, P. 2015, \mnras, 447, 646

\bibitem[{{Crocce} {et~al.}(2015){Crocce}, {Castander}, {Gazta{\~n}aga}, {Fosalba}, \& {Carretero}}]{crocce_2015}
{Crocce}, M., {Castander}, F.~J., {Gazta{\~n}aga}, E., {Fosalba}, P., \& {Carretero}, J. 2015, \mnras, 453, 1513

\bibitem[{{de Jong} {et~al.}(2013){de Jong}, {Verdoes Kleijn}, {Kuijken}, \& {Valentijn}}]{de2013}
{de Jong}, J. T.~A., {Verdoes Kleijn}, G.~A., {Kuijken}, K.~H., \& {Valentijn}, E.~A. 2013, Experimental Astronomy, 35, 25

\bibitem[{{Diemer}(2018)}]{diemer_2018}
{Diemer}, B. 2018, \apjs, 239, 35

\bibitem[{{Diemer}(2023)}]{diemer_2023}
{Diemer}, B. 2023, \mnras, 519, 3292

\bibitem[{{Diemer} \& {Joyce}(2019)}]{diemer_2019}
{Diemer}, B. \& {Joyce}, M. 2019, \apj, 871, 168

\bibitem[{{Driver} {et~al.}(2011){Driver}, {Hill}, {Kelvin}, {Robotham}, {Liske}, {Norberg}, {Baldry}, {Bamford}, {Hopkins}, {Loveday}, {Peacock}, {Andrae}, {Bland-Hawthorn}, {Brough}, {Brown}, {Cameron}, {Ching}, {Colless}, {Conselice}, {Croom}, {Cross}, {de Propris}, {Dye}, {Drinkwater}, {Ellis}, {Graham}, {Grootes}, {Gunawardhana}, {Jones}, {van Kampen}, {Maraston}, {Nichol}, {Parkinson}, {Phillipps}, {Pimbblet}, {Popescu}, {Prescott}, {Roseboom}, {Sadler}, {Sansom}, {Sharp}, {Smith}, {Taylor}, {Thomas}, {Tuffs}, {Wijesinghe}, {Dunne}, {Frenk}, {Jarvis}, {Madore}, {Meyer}, {Seibert}, {Staveley-Smith}, {Sutherland}, \& {Warren}}]{driver_2011}
{Driver}, S.~P., {Hill}, D.~T., {Kelvin}, L.~S., {et~al.} 2011, \mnras, 413, 971

\bibitem[{{Dvornik} {et~al.}(2017){Dvornik}, {Cacciato}, {Kuijken}, {Viola}, {Hoekstra}, {Nakajima}, {van Uitert}, {Brouwer}, {Choi}, {Erben}, {Fenech Conti}, {Farrow}, {Herbonnet}, {Heymans}, {Hildebrandt}, {Hopkins}, {McFarland}, {Norberg}, {Schneider}, {Sif{\'o}n}, {Valentijn}, \& {Wang}}]{dvornik_2017}
{Dvornik}, A., {Cacciato}, M., {Kuijken}, K., {et~al.} 2017, \mnras, 468, 3251

\bibitem[{{Edge} {et~al.}(2013){Edge}, {Sutherland}, {Kuijken}, {Driver}, {McMahon}, {Eales}, \& {Emerson}}]{edge_2013}
{Edge}, A., {Sutherland}, W., {Kuijken}, K., {et~al.} 2013, The Messenger, 154, 32

\bibitem[{{Fitzpatrick}(1999)}]{fitzp_1999}
{Fitzpatrick}, E.~L. 1999, \pasp, 111, 63

\bibitem[{{Flaugher} {et~al.}(2015){Flaugher}, {Diehl}, {Honscheid}, {Abbott}, {Alvarez}, {Angstadt}, {Annis}, {Antonik}, {Ballester}, {Beaufore}, {Bernstein}, {Bernstein}, {Bigelow}, {Bonati}, {Boprie}, {Brooks}, {Buckley-Geer}, {Campa}, {Cardiel-Sas}, {Castander}, {Castilla}, {Cease}, {Cela-Ruiz}, {Chappa}, {Chi}, {Cooper}, {da Costa}, {Dede}, {Derylo}, {DePoy}, {de Vicente}, {Doel}, {Drlica-Wagner}, {Eiting}, {Elliott}, {Emes}, {Estrada}, {Fausti Neto}, {Finley}, {Flores}, {Frieman}, {Gerdes}, {Gladders}, {Gregory}, {Gutierrez}, {Hao}, {Holland}, {Holm}, {Huffman}, {Jackson}, {James}, {Jonas}, {Karcher}, {Karliner}, {Kent}, {Kessler}, {Kozlovsky}, {Kron}, {Kubik}, {Kuehn}, {Kuhlmann}, {Kuk}, {Lahav}, {Lathrop}, {Lee}, {Levi}, {Lewis}, {Li}, {Mandrichenko}, {Marshall}, {Martinez}, {Merritt}, {Miquel}, {Mu{\~n}oz}, {Neilsen}, {Nichol}, {Nord}, {Ogando}, {Olsen}, {Palaio}, {Patton}, {Peoples}, {Plazas}, {Rauch}, {Reil}, {Rheault}, {Roe}, {Rogers}, {Roodman}, {Sanchez}, {Scarpine}, {Schindler}, {Schmidt},
  {Schmitt}, {Schubnell}, {Schultz}, {Schurter}, {Scott}, {Serrano}, {Shaw}, {Smith}, {Soares-Santos}, {Stefanik}, {Stuermer}, {Suchyta}, {Sypniewski}, {Tarle}, {Thaler}, {Tighe}, {Tran}, {Tucker}, {Walker}, {Wang}, {Watson}, {Weaverdyck}, {Wester}, {Woods}, {Yanny}, \& {DES Collaboration}}]{flaugher_2015}
{Flaugher}, B., {Diehl}, H.~T., {Honscheid}, K., {et~al.} 2015, \aj, 150, 150

\bibitem[{{Foreman-Mackey} {et~al.}(2013){Foreman-Mackey}, {Hogg}, {Lang}, \& {Goodman}}]{emcee}
{Foreman-Mackey}, D., {Hogg}, D.~W., {Lang}, D., \& {Goodman}, J. 2013, \pasp, 125, 306

\bibitem[{{Fosalba} {et~al.}(2015{\natexlab{a}}){Fosalba}, {Crocce}, {Gazta{\~n}aga}, \& {Castander}}]{fosalba_2015a}
{Fosalba}, P., {Crocce}, M., {Gazta{\~n}aga}, E., \& {Castander}, F.~J. 2015{\natexlab{a}}, \mnras, 448, 2987

\bibitem[{{Fosalba} {et~al.}(2015{\natexlab{b}}){Fosalba}, {Gazta{\~n}aga}, {Castander}, \& {Crocce}}]{fosalba_2015b}
{Fosalba}, P., {Gazta{\~n}aga}, E., {Castander}, F.~J., \& {Crocce}, M. 2015{\natexlab{b}}, \mnras, 447, 1319

\bibitem[{{Garcia-Fernandez} {et~al.}(2018){Garcia-Fernandez}, {Sanchez}, {Sevilla-Noarbe}, {Suchyta}, {Huff}, {Gaztanaga}, {Aleksi{\'c}}, {}, {Ponce}, {Castander}, {Hoyle}, {Abbott}, {Abdalla}, {Allam}, {Annis}, {Benoit-L{\'e}vy}, {Bernstein}, {Bertin}, {Brooks}, {Buckley-Geer}, {Burke}, {Carnero Rosell}, {Carrasco Kind}, {Carretero}, {Crocce}, {Cunha}, {D'Andrea}, {da Costa}, {DePoy}, {Desai}, {Diehl}, {Eifler}, {Evrard}, {Fernandez}, {Flaugher}, {Fosalba}, {Frieman}, {Garc{\'\i}a-Bellido}, {Gerdes}, {Giannantonio}, {Gruen}, {Gruendl}, {Gschwend}, {Gutierrez}, {James}, {Jarvis}, {Kirk}, {Krause}, {Kuehn}, {Kuropatkin}, {Lahav}, {Lima}, {MacCrann}, {Maia}, {March}, {Marshall}, {Melchior}, {Miquel}, {Mohr}, {Plazas}, {Romer}, {Roodman}, {Rykoff}, {Scarpine}, {Schubnell}, {Smith}, {Soares-Santos}, {Sobreira}, {Tarle}, {Thomas}, {Walker}, {Wester}, \& {DES Collaboration}}]{garcia_2018}
{Garcia-Fernandez}, M., {Sanchez}, E., {Sevilla-Noarbe}, I., {et~al.} 2018, \mnras, 476, 1071

\bibitem[{{Hartlap} {et~al.}(2007){Hartlap}, {Simon}, \& {Schneider}}]{hartlap_cov}
{Hartlap}, J., {Simon}, P., \& {Schneider}, P. 2007, \aap, 464, 399

\bibitem[{Hildebrandt {et~al.}(2021)Hildebrandt, van~den Busch, Wright, Blake, Joachimi, Kuijken, Tr{\"o}ster, Asgari, Bilicki, de~Jong, {et~al.}}]{hildebrandt2021kids}
Hildebrandt, H., van~den Busch, J., Wright, A., {et~al.} 2021, Astronomy \& Astrophysics, 647, A124

\bibitem[{{Jarvis} {et~al.}(2004){Jarvis}, {Bernstein}, \& {Jain}}]{treecorr}
{Jarvis}, M., {Bernstein}, \& {Jain}, B. 2004, \mnras, 352, 338

\bibitem[{{Kuijken}(2011)}]{kuijken_2011}
{Kuijken}, K. 2011, The Messenger, 146, 8

\bibitem[{{Kuijken} {et~al.}(2019){Kuijken}, {Heymans}, {Dvornik}, {Hildebrandt}, {de Jong}, {Wright}, {Erben}, {Bilicki}, {Giblin}, {Shan}, {Getman}, {Grado}, {Hoekstra}, {Miller}, {Napolitano}, {Paolilo}, {Radovich}, {Schneider}, {Sutherland}, {Tewes}, {Tortora}, {Valentijn}, \& {Verdoes Kleijn}}]{kuijken_2019}
{Kuijken}, K., {Heymans}, C., {Dvornik}, A., {et~al.} 2019, \aap, 625, A2

\bibitem[{{Kuijken} {et~al.}(2015){Kuijken}, {Heymans}, {Hildebrandt}, {Nakajima}, {Erben}, {de Jong}, {Viola}, {Choi}, {Hoekstra}, {Miller}, {van Uitert}, {Amon}, {Blake}, {Brouwer}, {Buddendiek}, {Conti}, {Eriksen}, {Grado}, {Harnois-D{\'e}raps}, {Helmich}, {Herbonnet}, {Irisarri}, {Kitching}, {Klaes}, {La Barbera}, {Napolitano}, {Radovich}, {Schneider}, {Sif{\'o}n}, {Sikkema}, {Simon}, {Tudorica}, {Valentijn}, {Verdoes Kleijn}, \& {van Waerbeke}}]{kuijken_2015}
{Kuijken}, K., {Heymans}, C., {Hildebrandt}, H., {et~al.} 2015, \mnras, 454, 3500

\bibitem[{{M{\'e}nard} {et~al.}(2003){M{\'e}nard}, {Hamana}, {Bartelmann}, \& {Yoshida}}]{menard_2003}
{M{\'e}nard}, B., {Hamana}, T., {Bartelmann}, M., \& {Yoshida}, N. 2003, \aap, 403, 817

\bibitem[{{M{\'e}nard} {et~al.}(2010){M{\'e}nard}, {Scranton}, {Fukugita}, \& {Richards}}]{menard_2010}
{M{\'e}nard}, B., {Scranton}, R., {Fukugita}, M., \& {Richards}, G. 2010, \mnras, 405, 1025

\bibitem[{{Navarro} {et~al.}(1997){Navarro}, {Frenk}, \& {White}}]{nfw_97}
{Navarro}, J.~F., {Frenk}, C.~S., \& {White}, S. D.~M. 1997, \apj, 490, 493

\bibitem[{{Peek} {et~al.}(2015){Peek}, {M{\'e}nard}, \& {Corrales}}]{peek_2015}
{Peek}, J.~E.~G., {M{\'e}nard}, B., \& {Corrales}, L. 2015, \apj, 813, 7

\bibitem[{{Planck Collaboration} {et~al.}(2020){Planck Collaboration}, {Aghanim}, {Akrami}, {Ashdown}, {Aumont}, {Baccigalupi}, {Ballardini}, {Banday}, {Barreiro}, {Bartolo}, {Basak}, {Battye}, {Benabed}, {Bernard}, {Bersanelli}, {Bielewicz}, {Bock}, {Bond}, {Borrill}, {Bouchet}, {Boulanger}, {Bucher}, {Burigana}, {Butler}, {Calabrese}, {Cardoso}, {Carron}, {Challinor}, {Chiang}, {Chluba}, {Colombo}, {Combet}, {Contreras}, {Crill}, {Cuttaia}, {de Bernardis}, {de Zotti}, {Delabrouille}, {Delouis}, {Di Valentino}, {Diego}, {Dor{\'e}}, {Douspis}, {Ducout}, {Dupac}, {Dusini}, {Efstathiou}, {Elsner}, {En{\ss}lin}, {Eriksen}, {Fantaye}, {Farhang}, {Fergusson}, {Fernandez-Cobos}, {Finelli}, {Forastieri}, {Frailis}, {Fraisse}, {Franceschi}, {Frolov}, {Galeotta}, {Galli}, {Ganga}, {G{\'e}nova-Santos}, {Gerbino}, {Ghosh}, {Gonz{\'a}lez-Nuevo}, {G{\'o}rski}, {Gratton}, {Gruppuso}, {Gudmundsson}, {Hamann}, {Handley}, {Hansen}, {Herranz}, {Hildebrandt}, {Hivon}, {Huang}, {Jaffe}, {Jones}, {Karakci}, {Keih{\"a}nen},
  {Keskitalo}, {Kiiveri}, {Kim}, {Kisner}, {Knox}, {Krachmalnicoff}, {Kunz}, {Kurki-Suonio}, {Lagache}, {Lamarre}, {Lasenby}, {Lattanzi}, {Lawrence}, {Le Jeune}, {Lemos}, {Lesgourgues}, {Levrier}, {Lewis}, {Liguori}, {Lilje}, {Lilley}, {Lindholm}, {L{\'o}pez-Caniego}, {Lubin}, {Ma}, {Mac{\'\i}as-P{\'e}rez}, {Maggio}, {Maino}, {Mandolesi}, {Mangilli}, {Marcos-Caballero}, {Maris}, {Martin}, {Martinelli}, {Mart{\'\i}nez-Gonz{\'a}lez}, {Matarrese}, {Mauri}, {McEwen}, {Meinhold}, {Melchiorri}, {Mennella}, {Migliaccio}, {Millea}, {Mitra}, {Miville-Desch{\^e}nes}, {Molinari}, {Montier}, {Morgante}, {Moss}, {Natoli}, {N{\o}rgaard-Nielsen}, {Pagano}, {Paoletti}, {Partridge}, {Patanchon}, {Peiris}, {Perrotta}, {Pettorino}, {Piacentini}, {Polastri}, {Polenta}, {Puget}, {Rachen}, {Reinecke}, {Remazeilles}, {Renzi}, {Rocha}, {Rosset}, {Roudier}, {Rubi{\~n}o-Mart{\'\i}n}, {Ruiz-Granados}, {Salvati}, {Sandri}, {Savelainen}, {Scott}, {Shellard}, {Sirignano}, {Sirri}, {Spencer}, {Sunyaev}, {Suur-Uski}, {Tauber}, {Tavagnacco},
  {Tenti}, {Toffolatti}, {Tomasi}, {Trombetti}, {Valenziano}, {Valiviita}, {Van Tent}, {Vibert}, {Vielva}, {Villa}, {Vittorio}, {Wandelt}, {Wehus}, {White}, {White}, {Zacchei}, \& {Zonca}}]{planck_2018}
{Planck Collaboration}, {Aghanim}, N., {Akrami}, Y., {et~al.} 2020, \aap, 641, A6

\bibitem[{{Sadeh} {et~al.}(2016){Sadeh}, {Abdalla}, \& {Lahav}}]{sadeh_2016}
{Sadeh}, I., {Abdalla}, F.~B., \& {Lahav}, O. 2016, \pasp, 128, 104502

\bibitem[{{Singh} {et~al.}(2017){Singh}, {Mandelbaum}, {Seljak}, {Slosar}, \& {Vazquez Gonzalez}}]{singh_2017}
{Singh}, S., {Mandelbaum}, R., {Seljak}, U., {Slosar}, A., \& {Vazquez Gonzalez}, J. 2017, \mnras, 471, 3827

\bibitem[{{Tinker} {et~al.}(2010){Tinker}, {Robertson}, {Kravtsov}, {Klypin}, {Warren}, {Yepes}, \& {Gottl{\"o}ber}}]{tinker_2010}
{Tinker}, J.~L., {Robertson}, B.~E., {Kravtsov}, A.~V., {et~al.} 2010, \apj, 724, 878

\bibitem[{{Triani} {et~al.}(2020){Triani}, {Sinha}, {Croton}, {Pacifici}, \& {Dwek}}]{triani_2020}
{Triani}, D.~P., {Sinha}, M., {Croton}, D.~J., {Pacifici}, C., \& {Dwek}, E. 2020, \mnras, 493, 2490

\bibitem[{{van den Busch} {et~al.}(2020){van den Busch}, {Hildebrandt}, {Wright}, {Morrison}, {Blake}, {Joachimi}, {Erben}, {Heymans}, {Kuijken}, \& {Taylor}}]{jlvdb_2020}
{van den Busch}, J.~L., {Hildebrandt}, H., {Wright}, A.~H., {et~al.} 2020, \aap, 642, A200

\bibitem[{{Weingartner} \& {Draine}(2001)}]{w_draine_2001}
{Weingartner}, J.~C. \& {Draine}, B.~T. 2001, \apj, 548, 296

\bibitem[{{Wright} {et~al.}(2024){Wright}, {Kuijken}, {Hildebrandt}, {Radovich}, {Bilicki}, {Dvornik}, {Getman}, {Heymans}, {Hoekstra}, {Li}, {Miller}, {Napolitano}, {Xia}, {Asgari}, {Brescia}, {Buddelmeijer}, {Burger}, {Castignani}, {Cavuoti}, {de Jong}, {Edge}, {Giblin}, {Giocoli}, {Harnois-D{\'e}raps}, {Jalan}, {Joachimi}, {John William}, {Joudaki}, {Kannawadi}, {Kaur}, {La Barbera}, {Linke}, {Mahony}, {Maturi}, {Moscardini}, {Nakoneczny}, {Paolillo}, {Porth}, {Puddu}, {Reischke}, {Schneider}, {Sereno}, {Shan}, {Sif{\'o}n}, {St{\"o}lzner}, {Tr{\"o}ster}, {Valentijn}, {van den Busch}, {Verdoes Kleijn}, {Wittje}, {Yan}, {Yao}, {Yoon}, \& {Zhang}}]{wright_2024}
{Wright}, A.~H., {Kuijken}, K., {Hildebrandt}, H., {et~al.} 2024, \aap, 686, A170

\bibitem[{{York} {et~al.}(2000){York}, {Adelman}, {Anderson}, {Anderson}, {Annis}, {Bahcall}, {Bakken}, {Barkhouser}, {Bastian}, {Berman}, {Boroski}, {Bracker}, {Briegel}, {Briggs}, {Brinkmann}, {Brunner}, {Burles}, {Carey}, {Carr}, {Castander}, {Chen}, {Colestock}, {Connolly}, {Crocker}, {Csabai}, {Czarapata}, {Davis}, {Doi}, {Dombeck}, {Eisenstein}, {Ellman}, {Elms}, {Evans}, {Fan}, {Federwitz}, {Fiscelli}, {Friedman}, {Frieman}, {Fukugita}, {Gillespie}, {Gunn}, {Gurbani}, {de Haas}, {Haldeman}, {Harris}, {Hayes}, {Heckman}, {Hennessy}, {Hindsley}, {Holm}, {Holmgren}, {Huang}, {Hull}, {Husby}, {Ichikawa}, {Ichikawa}, {Ivezi{\'c}}, {Kent}, {Kim}, {Kinney}, {Klaene}, {Kleinman}, {Kleinman}, {Knapp}, {Korienek}, {Kron}, {Kunszt}, {Lamb}, {Lee}, {Leger}, {Limmongkol}, {Lindenmeyer}, {Long}, {Loomis}, {Loveday}, {Lucinio}, {Lupton}, {MacKinnon}, {Mannery}, {Mantsch}, {Margon}, {McGehee}, {McKay}, {Meiksin}, {Merelli}, {Monet}, {Munn}, {Narayanan}, {Nash}, {Neilsen}, {Neswold}, {Newberg}, {Nichol}, {Nicinski},
  {Nonino}, {Okada}, {Okamura}, {Ostriker}, {Owen}, {Pauls}, {Peoples}, {Peterson}, {Petravick}, {Pier}, {Pope}, {Pordes}, {Prosapio}, {Rechenmacher}, {Quinn}, {Richards}, {Richmond}, {Rivetta}, {Rockosi}, {Ruthmansdorfer}, {Sandford}, {Schlegel}, {Schneider}, {Sekiguchi}, {Sergey}, {Shimasaku}, {Siegmund}, {Smee}, {Smith}, {Snedden}, {Stone}, {Stoughton}, {Strauss}, {Stubbs}, {SubbaRao}, {Szalay}, {Szapudi}, {Szokoly}, {Thakar}, {Tremonti}, {Tucker}, {Uomoto}, {Vanden Berk}, {Vogeley}, {Waddell}, {Wang}, {Watanabe}, {Weinberg}, {Yanny}, {Yasuda}, \& {SDSS Collaboration}}]{york_2000}
{York}, D.~G., {Adelman}, J., {Anderson}, John~E., J., {et~al.} 2000, \aj, 120, 1579

\end{thebibliography}

%%%%%%%%%%
\appendix

\section{Magnitude shift in other bands}
\label{app:intrinsic}
\begin{figure*}
    \centering
    \includegraphics[width=0.96\textwidth]{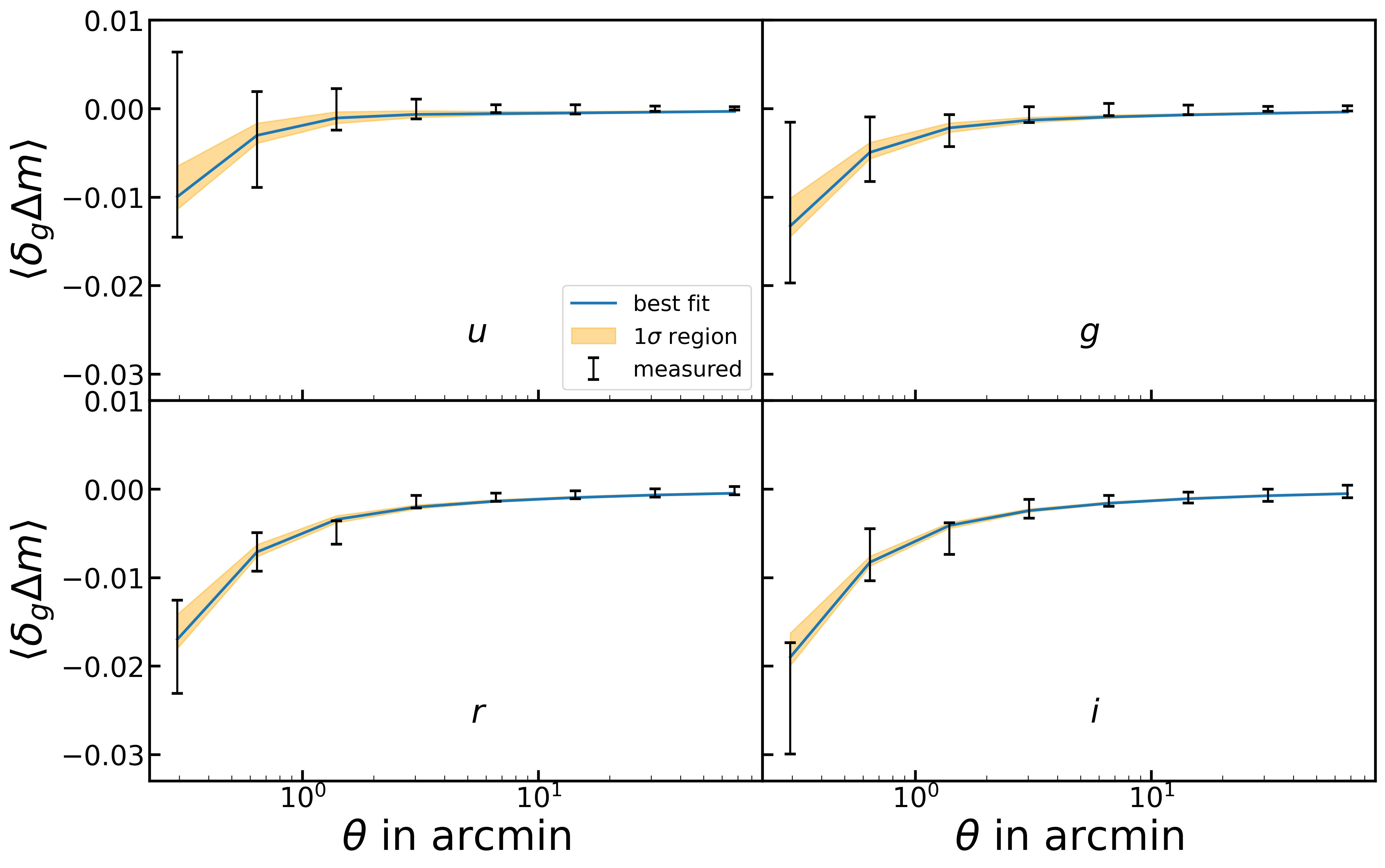}
    \caption{Magnitude shift measurements in $ugri$ bands and MCMC results.
Measured magnitude shift signals (black bars) in each optical band compared to the best-fit model, which is the blue solid line resulted from the MCMC analysis. We also show the $1\sigma$ posterior region, which is indicated by the yellow shaded region.}
    \label{fig:mice_magn_ugri}
\end{figure*}

In this section, we present the magnitude shift measurements for the $ugri$ bands using MICE2 galaxies, where we include realistic observational effects: noisy magnitudes, photometric redshifts, and band-specific magnitude cuts applied to the source sample. These measurements correspond to the same setup as used in our main analysis.
\\
In Fig.~\ref{fig:mice_magn_ugri} we show the measured data points with error bars, overlaid with the best-fit model prediction obtained from our joint MCMC fit to the GGL and multi-band magnitude shift signals. The shaded regions represent the $1\sigma$ posterior credible intervals of the fit. This figure demonstrates that our modelling framework accurately reproduces the magnitude shift signals in all four photometric bands. The best-fit model traces the data well within uncertainties, confirming the robustness of our signal extraction and the reliability of the MCMC inference.

\clearpage

\section{Full parameter constraints from MCMC analysis of KiDS}
\label{app:full_mcmc_constraints}

\begin{figure*}
    \centering
    \includegraphics[width=\columnwidth]{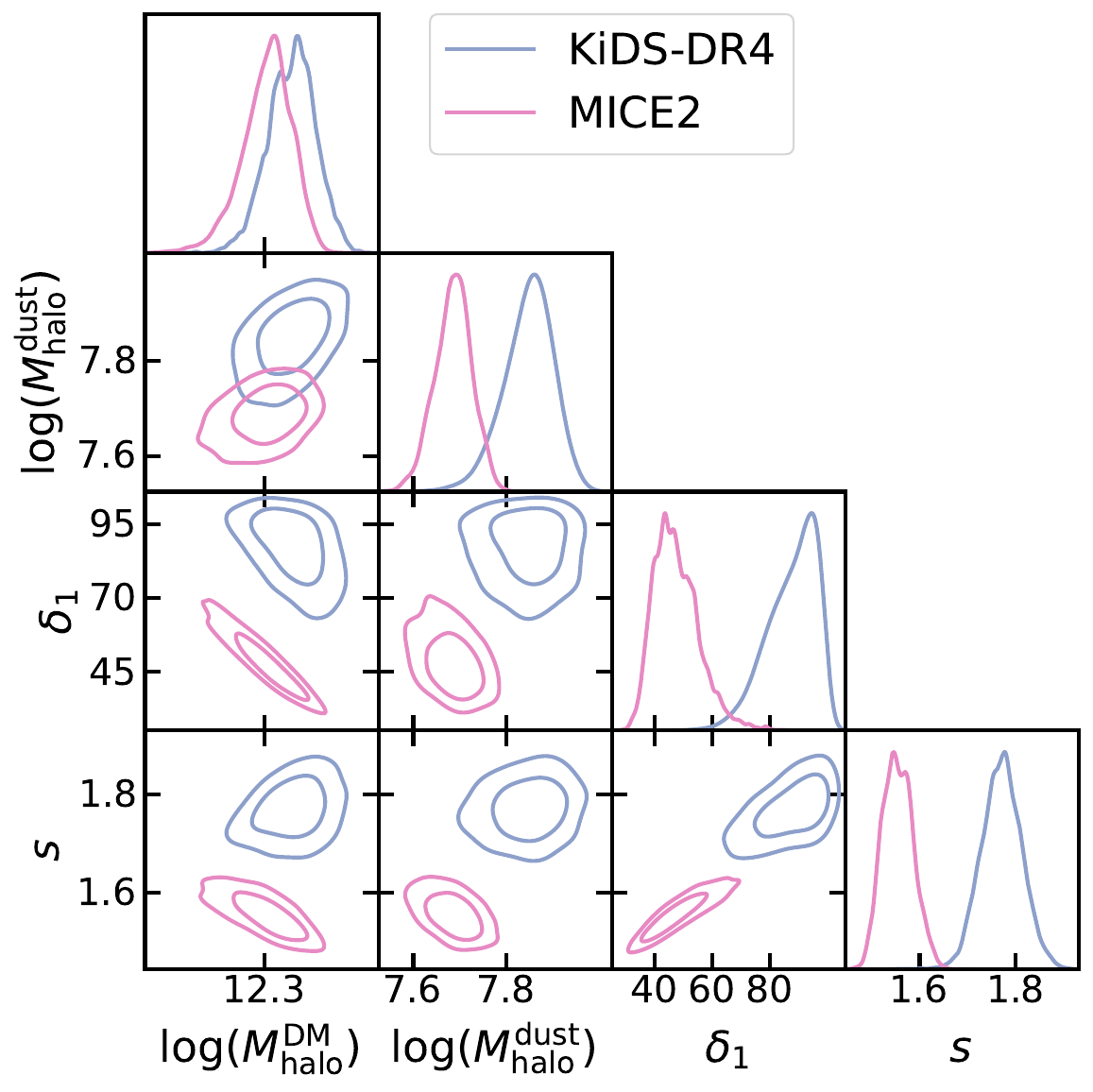}
    \caption{Full set of parameter constraints obtained from the MCMC analysis of KiDS galaxies. We present the joint analysis of GGL and magnification in four optical bands.}
    \label{fig:app_8}
\end{figure*}

In this section, we present the full set of parameter constraints obtained from the MCMC analysis of MICE2 and KiDS galaxies in Fig.~\ref{fig:app_8}. We show the results of the joint analysis of GGL signal and magnitude shift signals in $ugri$ bands.
In addition to the halo mass and dust mass, the other two parameters—$\delta_1$ and $s$—are also included here for completeness. All parameters are well-constrained within the explored parameter space. However, the parameter $\delta_1$ exhibits a distribution that leans towards the upper boundary of its prior, with a peak near 90 and the upper limit set at 100. While this indicates that the parameter is approaching the edge of the allowed range, the peak suggests it is still constrained effectively without fully saturating the prior.

\clearpage
\section{Effects of using multi-band data on our measurements}
\label{app:kids_multiband}
\begin{figure}
    \centering
    \includegraphics[width=\columnwidth]{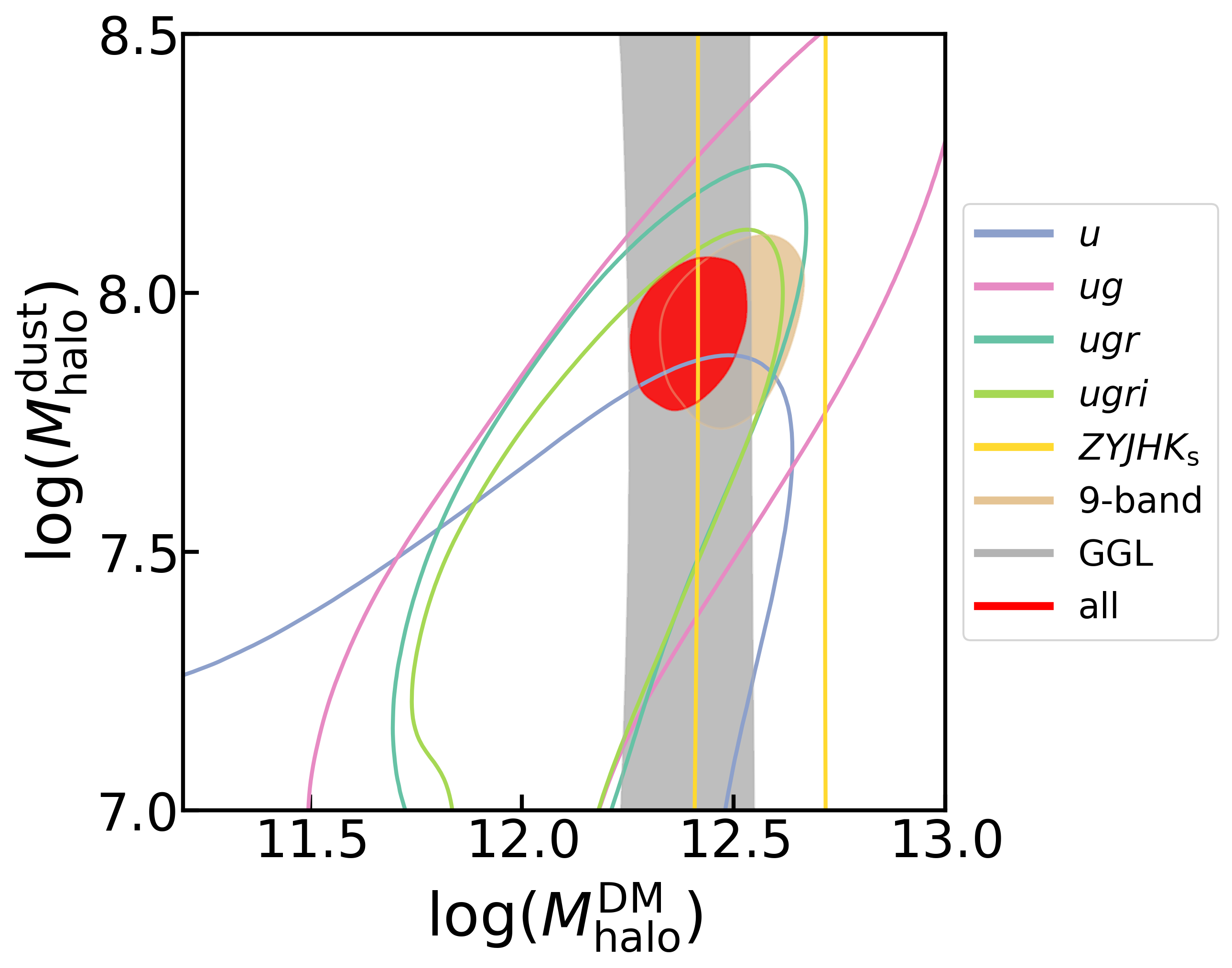}
    \caption{$2\sigma$ contours representing the estimations of halo mass and dust mass derived from magnitude shift in various optical and NIR bands, the GGL signal, and their joint analyses. The filled brown contour corresponds to the 9-band magnitude shift analysis, the grey contour to the GGL signal analysis, and the red contour to the combined analysis of 9-band magnitude shift and GGL signals.}
    \label{fig:app_6}
\end{figure}

As previously mentioned, the GGL signal is crucial for constraining the halo mass, and by extension, indirectly constraining the dust mass. Multi-band photometry is also essential for extracting extinction information, as dust impacts the magnitude shift signal differently across wavelengths. In Fig.~\ref{fig:app_6}, the $2\sigma$ contours of halo and dust mass estimations are shown using magnitude shift measurements across different bands alongside the GGL measurement. We observe that the $u$-band magnitude shift alone provides almost no constraint on halo or dust mass, while the contours tighten as more bands are included. This is likely due to the $u$-band being the shallowest of the nine bands. Additionally, the magnitude shift signals in NIR bands offer no constraint on the dust mass but help further constrain the halo mass. The fact that magnitude shift signals the NIR bands provide no constraints on the dust mass serves as a useful test for evaluating the systematic effects in our measurement pipeline.
These findings reaffirm that while extinction effects are subtle, multi-band photometry is necessary to extract them. Optical bands probe extinction, while NIR bands help tighten constraints on halo mass.

\begin{figure}
    \centering
    \includegraphics[width=\columnwidth]{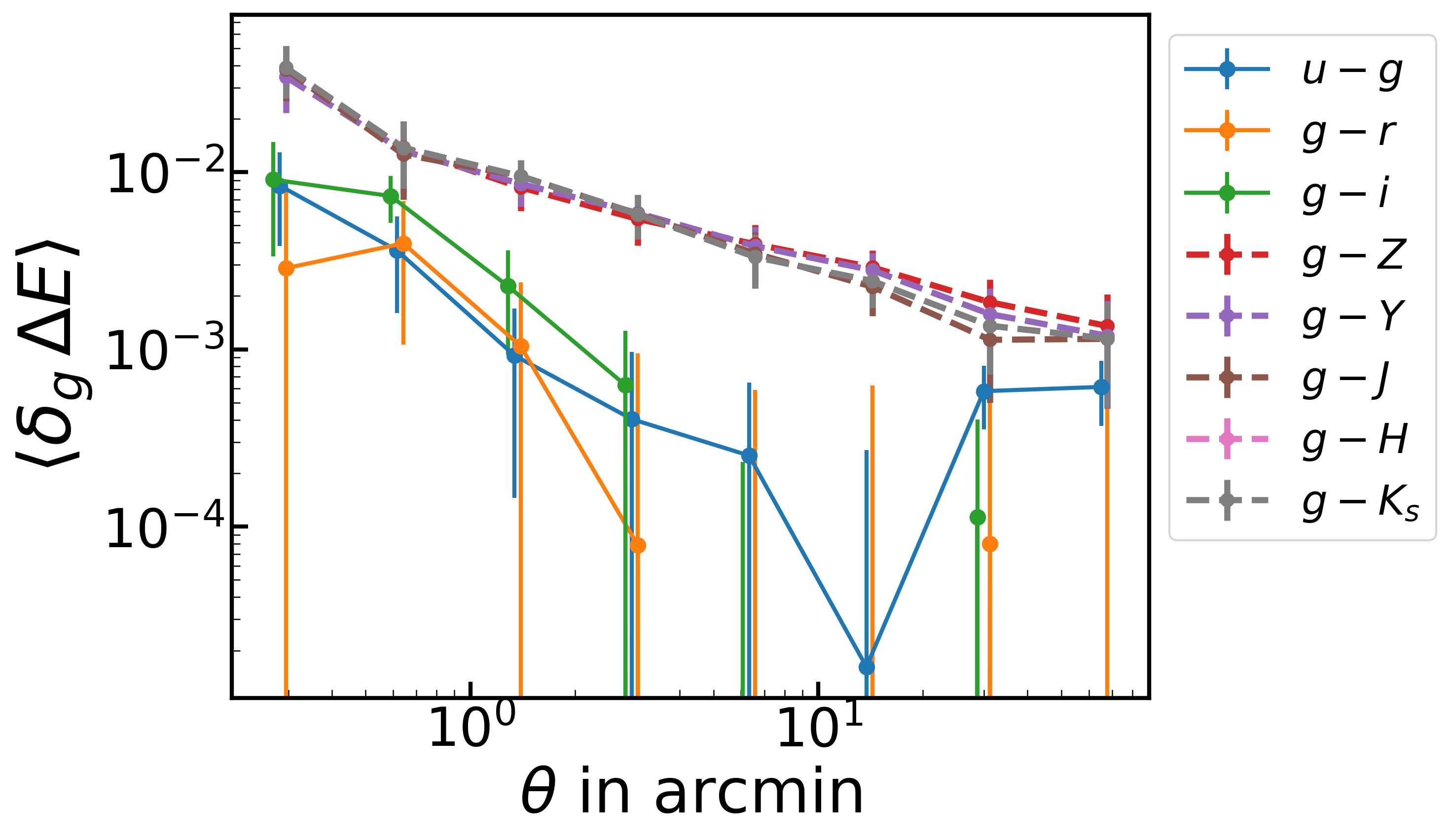}
    \caption{Comparison of optical and NIR colour excess signals. Colour excess measurements using $g - X_{\rm NIR}$  achieve higher S/N than purely optical colours, owing to the longer wavelength baseline.}
    \label{fig:app_7}
\end{figure}

Furthermore, we compare colour excess measurements involving both optical and NIR bands, to investigate the wavelength dependence of the dust extinction signal. In Fig.~\ref{fig:app_7}, we present a selection of $g-X$ colours, where $X$ spans from the $u$-band to the $K_\mathrm{s}$-band. For the $u$-band, we instead plot $u – g$ for consistency in visual interpretation.

We find that the NIR colour excess signals are all consistent with each other within uncertainties, as expected from the physical behaviour of dust. Photons in the NIR are less susceptible to scattering and absorption by dust, allowing these bands to act as an effective null test. This is a significant result, as it is the first time NIR magnitude shift signals have been used in this way in the literature.

In contrast, colour excess signals involving optical bands show stronger deviations, indicative of dust reddening. Among them, $u - g$ exhibits the lowest S/N due to the limited depth of the $u$-band, making the measurement noisier.

\section{Magnitude and Colour Distributions of Source Galaxies}
\label{app:colordist}
\begin{figure*}
    \centering
    \includegraphics[width=\columnwidth]{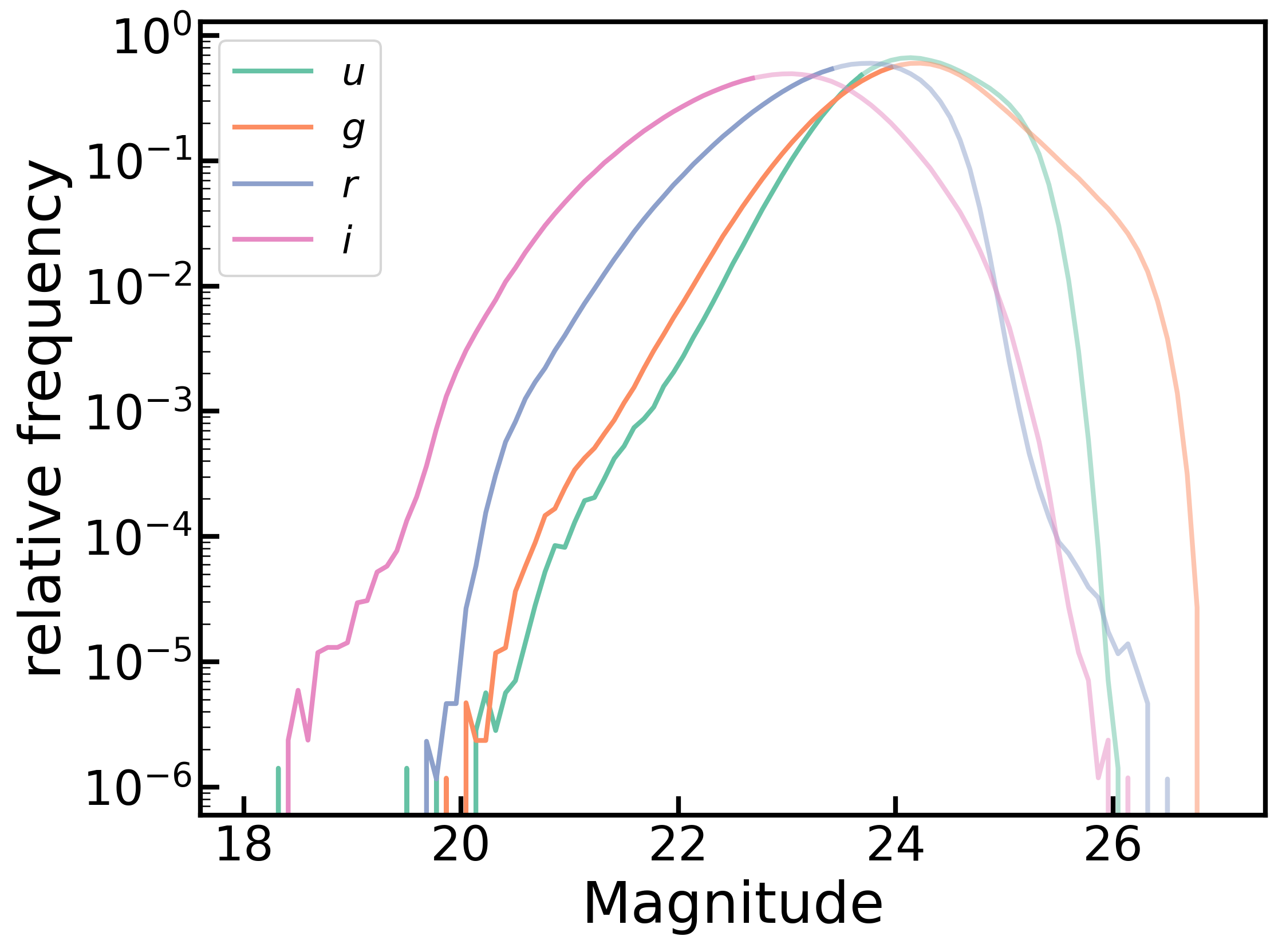}
    \includegraphics[width=\columnwidth]{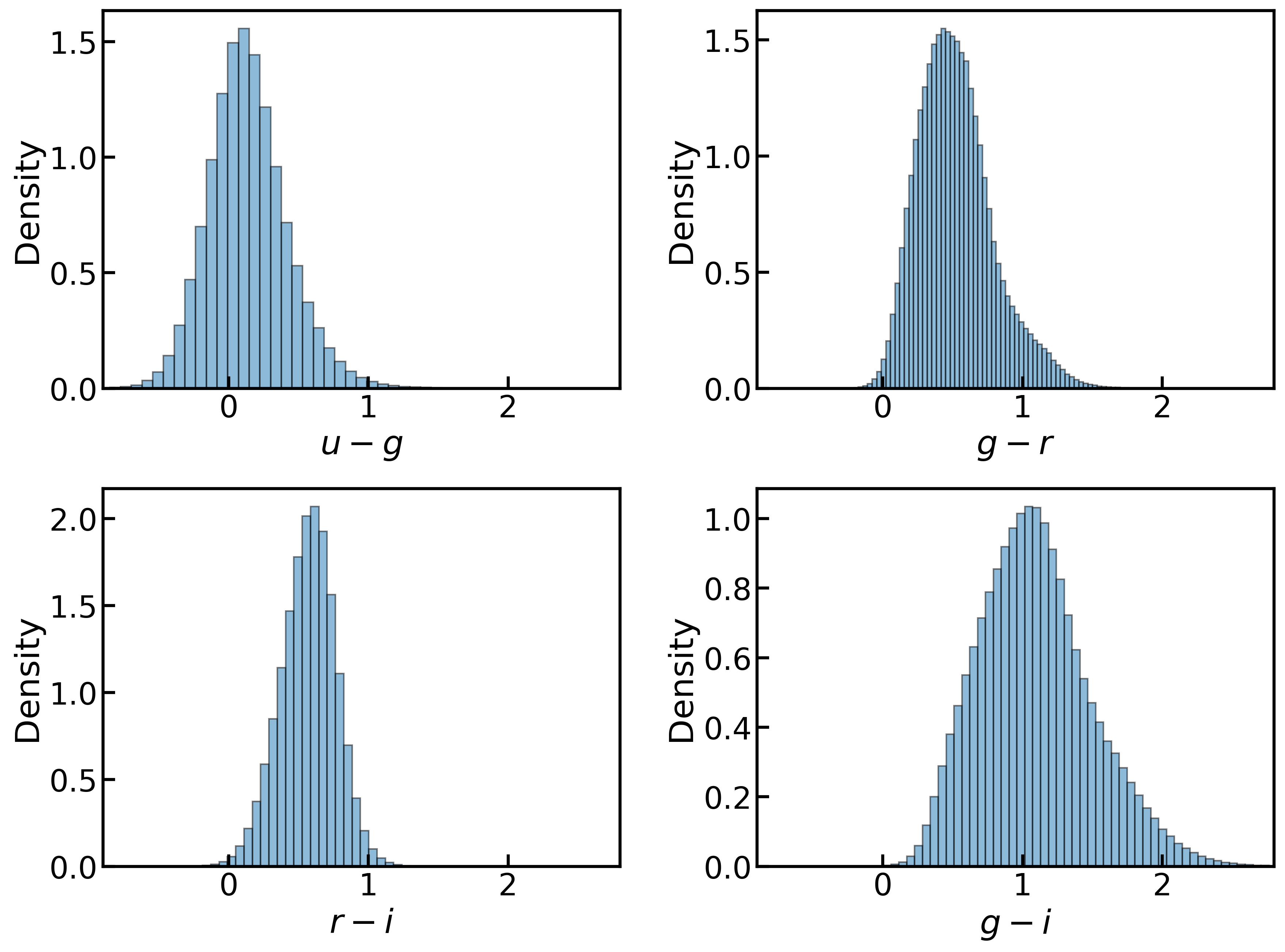}
    \caption{Magnitude and colour distributions of source galaxies in KiDS, on left and right hand side, respectively. Our distributions are broader, particularly in colour, which mostly explains the lower S/N. Quantitative widths are listed in Table \ref{tab:widths}.}
    \label{fig:magcolor_dists}
\end{figure*}

In Fig.~\ref{fig:magcolor_dists}, we present the optical magnitude and colour distributions of our source galaxies on the left and right panels, respectively. Compared to the analogous plots of \citet{menard_2010} (their Fig.2), we see that our colour distributions are broader. Table~\ref{tab:widths} summarises the widths of the colour distributions, quantified by both the standard deviation ($\sigma$) and the interquartile range (IQR), for the $u - g$, $g - r$, $r - i$, and $g - i$ colours. Our values are consistently larger than those of \citet{menard_2010}, with the $g - i$ colour showing almost a factor of three difference. These numbers for \citet{menard_2010} were obtained by digitising their published colour distributions.
\\
Moreover, our sample contains 362,064 lens galaxies and $\sim3.6$ million source galaxies, while \citet{menard_2010} used 24 million lenses and 85,704 quasars. This implies that, based on source and lens counts alone, we would expect a S/N approximately 0.8 times that of \citet{menard_2010}. Combined with the broader colour distributions in our sample, this roughly accounts for the observed factor of 2.6 difference in S/N of the $g - i$ colour excess signal . This analysis highlights that both the number of galaxy pairs and the intrinsic colour scatter are important in determining the S/N of the reddening measurements.

\begin{table}[h]
\centering
\caption{Widths of colour distributions for our sample and \citet{menard_2010}.}
\begin{tabular}{lcccc}
\hline
\hline
 & u-g & g-r & r-i & g-i \\
\hline
\textbf{This work} & & & & \\
$\sigma$ & 0.286 & 0.272 & 0.198 & 0.407 \\
IQR & 0.354 & 0.347 & 0.262 & 0.533 \\
\hline
\textbf{\citet{menard_2010}} & & & & \\
$\sigma$ & 0.152 & 0.144 & 0.115 & 0.147 \\
IQR & 0.208 & 0.177 & 0.170 & 0.197 \\
\hline
\end{tabular}
\label{tab:widths}
\end{table}

\end{document}